\documentclass[final,3p,times]{elsarticle}

\usepackage{amssymb}

\usepackage{tabularx}

\biboptions{compress}

\journal{Computers \& Mathematics with Applications}

\begin{document}

\thispagestyle{empty}
\noindent {\bf Preprint}\\
~\\
\noindent Hiroki Sayama, Irene Pestov, Jeffrey Schmidt, Benjamin James
Bush, Chun Wong, Junichi Yamanoi, and Thilo Gross, Modeling complex
systems with adaptive networks, {\em Computers and Mathematics with
  Applications,} 2013, http://dx.doi.org/10.1016/j.camwa.2012.12.005.\\

~\\
\noindent {\bf Disclaimer:} One of the authors carried out this
research on behalf of the Government of Canada, and as such the
copyright for that part of the paper belongs to the Crown, that is to
the Canadian Government. No provision exists for the transfer of any
such Crown copyright. Non-exclusive permission is granted to use,
reproduce and communicate the copyright article in any way you wish,
as long as its source is acknowledged.

\newpage

\setcounter{page}{1}

\begin{frontmatter}



\title{Modeling Complex Systems with Adaptive Networks}


\author[bing]{Hiroki Sayama\corref{cor}}
\cortext[cor]{Corresponding author}
\ead{sayama@binghamton.edu}
\author[drdc]{Irene Pestov}
\author[bing]{Jeffrey Schmidt}
\author[bing]{Benjamin James Bush}
\author[bing]{Chun Wong}
\author[chuo]{Junichi Yamanoi}
\author[bris]{Thilo Gross}

\address[bing]{Collective Dynamics of Complex Systems Research Group, Binghamton University, State University of New York, Binghamton, NY, USA}
\address[drdc]{Defence Research \& Development Canada, Centre for Operational Research and Analysis, Ottawa, Ontario}
\address[chuo]{Faculty of Policy Studies, Chuo University, Tokyo, Japan}
\address[bris]{Department of Engineering Mathematics, University of Bristol, Bristol, UK}

\begin{abstract}
Adaptive networks are a novel class of dynamical networks whose
topologies and states coevolve. Many real-world complex systems can be
modeled as adaptive networks, including social networks,
transportation networks, neural networks and biological networks. In
this paper, we introduce fundamental concepts and unique properties of
adaptive networks through a brief, non-comprehensive review of recent
literature on mathematical/computational modeling and analysis of such
networks. We also report our recent work on several applications of
computational adaptive network modeling and analysis to real-world
problems, including temporal development of search and rescue
operational networks, automated rule discovery from empirical network
evolution data, and cultural integration in corporate merger.
\end{abstract}


\end{frontmatter}



\section{Introduction}
\label{intro}

The rapidly growing research on complex networks has presented a new
approach to complex systems modeling and analysis
\cite{lit12,newman2003,lit3,pr-review2006,rmp-review2008}. It
addresses the self-organization of complex network structure and its
implications for system behavior, which holds significant
cross-disciplinary relevance to many fields of natural and social
sciences, particularly in today's highly networked society.

Interestingly, complex network research has historically addressed
either ``dynamics {\em on} networks'' or ``dynamics {\em of}
networks'' almost separately, without much consideration given to both
at the same time. In the former, ``dynamics {\em on} networks''
approach, the focus is on the state transition of nodes on a network
with a fixed topology and the trajectories of the system states in a
well-defined phase space
\cite{Pastor-Satorras2001,syncbook2003,sood2005,lit7,lit8,otto2007,nakao2010}. This
is a natural extension of traditional dynamical systems research to a
high-dimensional phase space with non-trivial interaction between
state variables. On the other hand, in the latter, ``dynamics {\em of}
networks'' approach, the focus is on the topological transformation of
a network and its effects on statistical properties of the entire
network \cite{price1965,lit1,lit10,lit13,lit14,krapivsky2005,lit15},
where a number of key concepts and techniques utilized are borrowed
from statistical physics and social network analysis.


When looking into real-world complex networks, however, one can find
many instances of networks whose states and topologies ``coevolve'',
i.e., they interact with each other and keep changing, often over the
same time scales, due to the system's own dynamics (Table
\ref{examples}). In these ``adaptive networks'', state transition of
each component and topological transformation of networks are deeply
coupled with each other, producing emergent behavior that would not be
seen in other forms of networks. Modeling and predicting
state-topology coevolution is now becoming well recognized as one of
the most significant challenges in complex network research
\cite{lit12,rmp-review2008,lit16,lit17}.

\begin{table}
\centering
\caption{Real-world examples of adaptive networks whose states and
  topologies interact with each other and coevolve.}
\newcolumntype{Y}{>{\raggedright\arraybackslash}X}
\begin{tabularx}{\textwidth}{Y|YYYYY}
\hline
\hline
{\bf Network} & {\bf Nodes} & {\bf Links} & {\bf Examples of node states} & {\bf Examples of node addition or removal} & {\bf Examples of topological changes} \\
\hline
{\em Organism} & Cells & Cell adhesions, intercellular communications & Gene/protein activities & Cell division, cell death & Cell migration \\
\hline
{\em Ecological community} & Species & Ecological relationships (predation, symbiosis, etc.) & Population, intraspecific diversities & Speciation, invasion, extinction & Changes in ecological relationships via adaptation \\
\hline
{\em Epidemiological network} & Individuals & Physical contacts & Pathologic states & Death, quarantine & Reduction of physical contacts \\
\hline
{\em Social network} & Individuals & Social relationships, conversations, collaborations & Socio-cultural states, political opinions, wealth & Entry to or withdrawal from community & Establishment or renouncement of relationships \\
\hline
\end{tabularx}
\label{examples}
\end{table}

In this paper, we introduce fundamental concepts and unique properties
of adaptive networks through a brief, non-comprehensive review of
recent literature on mathematical/computational modeling and analysis
of such networks. We also report our recent work on several
applications of computational adaptive network modeling and analysis
to real-world problems, including temporal development of search and
rescue operational networks, automated rule discovery from empirical
network evolution data, and cultural integration in corporate merger.

The rest of the paper is structured as follows. In the next section,
some of the recent literature is reviewed briefly to illustrate the
increasing attention to the field of adaptive networks. In Section
\ref{gna}, we introduce Generative Network Automata (GNA), a
theoretical framework for modeling adaptive network that we have
proposed. In Sections \ref{Opnet}--\ref{modeldiscovery}, we present
the aforementioned three examples of applications of adaptive network
modeling to study the dynamics of complex systems. The final section
summarizes and concludes the paper.

\section{Growing Literature on Adaptive Networks}
\label{literature}

Over the past decade, several mathematical/computational models of
state-topology coevolution in adaptive networks have been developed
and studied on various subjects, ranging from physical, biological to
social and engineered systems. In this section, we introduce a small
number of samples taken from the recent literature, categorizing them
into five major subjects of interest in the field.

\subsection{Self-Organized Criticality in Adaptive Neural Systems}

The present interest in adaptive networks was triggered by a paper
published by Bornholdt and Rohlf in 2000 \cite{lit18}. They built on
an observation by Christensen et al.~\cite{Christensen} who
investigated the dynamics of a simple dynamical model for extremal
optimization on complex networks.  In the penultimate paragraph of
their paper, Christensen et al.{} remarked that letting the structure
of the network coevolve with the dynamics on the network, leads to a
peculiar self-organization such that topological properties of the
network approached a critical point, where the dynamics on the network
changed qualitatively.

Inspired by Christensen et al., Bornholdt and Rohlf proposed a
different model in which the self-organization to the critical state
could be understood in greater detail. Their investigation showed that
the dynamical processes taking place on the network effectively
explored the network topology and thereby made topological information
available in every network node. This information then fed into the
local topological evolution and steered the dynamics toward the
critical state.  Thus a global self-organization is possible through
the interplay of two local processes.  Importantly, the paper of
Bornholdt and Rohlf demonstrated that this is not only the case in
rare, specifically engineered examples, but should be expected under
fairly general conditions.

Self-organized criticality is interesting because it can be argued
that every information processing system should be in a critical
state. It was therefore suspected that also the brain should reside in
a critical state \cite{chialvo}. The mechanism of Bornholdt and Rohlf
provided a plausible mechanism, explaining how criticality in the
brain could be achieved.  This ``criticality hypothesis"
\cite{Houghton} was subsequently supported by further models
\cite{Levina,Meisel} and laboratory experiments
\cite{Plenz,Kitzbichler}.
  
Today there is growing evidence that self-organized criticality is a
central process for brain functionality. Adaptive networks models
remain a major tool for understanding this process.  In the neural
context, understanding self-organized criticality in adaptive networks
is thus paving the way to new diagnostic tools and a deeper
understanding of neural disorders \cite{Meisel2}.  Furthermore,
understanding self-organized criticality in biological neural networks
is thought to hold the key to the artificial systems that can
self-organize to a state where they can process information.  This may
be enable the use of future nano-scale electronic components that are
too small to arrange precisely using photolitography, and thus have to
use adaptive principles to self-tune to a functional state after
quasi-random assembly.

\subsection{Epidemics on Adaptive Networks}

While adaptive self-organized criticality requires dynamics on
different time scales, other dynamical phenomena in adaptive networks
occur when topology and node states evolve simultaneously. The
resulting interplay has been investigated in detail in a class of
epidemiological models where the agents rewire their social contacts
in response to the epidemic state of other agents.

The first adaptive-network-based epidemic model was the adaptive
Susceptible-Infected-Susceptible (SIS) model studied by Gross et
al.~\cite{lit21}. By a so-called moment closure approximation, the
authors were able to compute transition points in the model
analytically. The main value of this work was to provide detailed
analytical insights into the emergence of system-level phenomena from
the node-level coevolution.  Today, the model remains a benchmark for
the performance of analytical approximations to adaptive networks
\cite{Gross,Marceau,Guerra,Nunes}.  Furthermore, it triggered a large
body subsequent investigations into the effect of social responses to
epidemics on disease propagation and vaccination strategies
\cite{Wang,lit22,Zanette,Volz}.  

Adaptive networks have produced significant implications for
real-world epidemiological practice, as they capture more realistic
dynamics of social networks where people tend to alter social
behaviors according to epidemiological states of their neighbors
\cite{funk2010}. For example, Epstein et al.~\cite{epstein} and Funk
et al.~\cite{Jansen} considered a spatial context for the influence of
human behavior in the outbreak of epidemics (although the former did
not explicitly use network models). Also, Shaw and Schwartz
\cite{shaw10} recently showed that vaccine control of a disease is
significantly more effective in adaptive social networks than in
static ones, because of individuals' adaptive behavioral responses to
the vaccine application.

\subsection{Adaptive Opinion Formation and Collective Behavior}

Another active direction in adaptive networks research focuses on
models of collective opinion formation.  These models describe the
diffusion of competing opinions through a networked population, where
agents can modify their contacts depending on the opinions held by
their neighbors.  Two similar pioneering models in this direction were
published by Holme and Newman \cite{lit24} and Zanette and Gil
\cite{ZanetteGil} in 2006.

A central question in opinion formation is whether the coevolutionary
dynamics will eventually lead to consensus or to a fragmentation
splitting the population into two disconnected camps. The transition
points between these two long-term outcomes is known as the
fragmentation transition.  The simplest and best-understood model
exhibiting the fragmentation transition is the adaptive voter model
\cite{lit28}.  A detailed understanding of the fragmentation
transition in this model was gained through the work of Vazquez et
al.~\cite{Vazquez} and the independent parallel study of Kimura and
Hayakawa \cite{Kimura}.

Although the adaptive voter model is similar to the adaptive SIS
model, analytical tools that perform well for the SIS model yield poor
results for the voter model \cite{Kimura}.  Nevertheless, the
transition point can be computed analytically, using a different
approach \cite{Boehme1,Boehme2}.

It was sometimes criticized that mathematical models of opinion
formation fall short of the complexity of opinion formation processes
in the real world, and that hence no connection to real-world
observations and experiments can be made. However, Centola et
al.~\cite{centola2007} studied agent-based adaptive network models of
more realistic cultural drift and dissemination processes, finding
similar dynamics including the fragmentation transition. Centola also
experimentally examined how social network structures interact with
human behaviors \cite{centola2010,centola2011}. More recently, the
works 
of Huepe et al.~\cite{huepe2011} and Couzin et al.~\cite{couzin2011}
showed that voter-like models can be used to understand the dynamics
of decision making in the collective motion of swarms of locusts
\cite{huepe2011} and schools of fish \cite{couzin2011}.  Their studies
demonstrated that analytically tractable adaptive network models could
predict the result of laboratory experiments.

\subsection{Social Games on Adaptive Networks}
Besides opinion formation, also other types of social dynamics have been 
investigated on adaptive networks. In particular, many adaptive extensions
of classical game theoretical models have been proposed. 

Three early works that appeared already in 2000 are a study of the
minority game on adaptive networks by Paczuski, Bassler, and Corral
\cite{Paczuski}, an exploration of various coordination and
cooperation games by Skyrms and Pemantle \cite{Skyrms}, and a study of
the Prisoner's dilemma by Zimmermann et al.~\cite{Zimmermann}.
Another influential work is a paper by by Bornholdt and Ebel, which
remains unfinished but is available as a preprint
\cite{BornholdtEbel}.

These papers above triggered a large number of subsequent work that
explored how coevolutionary dynamics affects the evolution of
cooperation in adaptive networks.  Notable examples are the work of
Pacheco et al.~\cite{lit25} and van Segbroeck et al.~\cite{Segbroeck}
who demonstrated clearly that coevolution can lead to increased levels
of cooperation; Poncela et al.~\cite{Poncela}, who showed that
coevolutionary dynamics can facilitate cooperation not only by
building up beneficial structures, but through the dynamics of growth
itself; and Zschaler et~al.~\cite{Zschaler}, who identified an
unconventional dynamical mechanism leading to full cooperation.

Research in adaptive networks also gave rise to a different class of
games, where agents do not aim to optimize some abstract payoff, but
struggle for an advantageous position in the network. The earliest
example of these adaptive network formation games is perhaps the paper
of Bala and Goyal \cite{BalaGoyal} which was published in
2001. Another early paper is Holme and Ghoshal's model \cite{lit23},
where the nodes tried to maximize their centralities by adaptively
changing their links based on locally available information, without
paying too much costs (i.e., maintaining too many connections). The
resulting time evolution of the network was highly nontrivial,
involving a cascade of strategic and topological changes, leading to a
network state that was close to the transition between well-connected
and fragmented states. A recent work by Do et al.~\cite{DoPOC}
presents an analytical investigation of network formation and
cooperation on an adaptive weighted network.

\subsection{Organizational Dynamics as Adaptive Networks}
Applications of adaptive networks do not stop at abstract social
models like those reviewed above. One of the latest application areas
of adaptive networks is the computational modeling of complex
organizational behavior, including the evolution of organizational
networks, information/knowledge/culture sharing and trust formation
within a group or corporation. Studies on organizational network
structures actually have several decades of history (including the
well-known structural holes argument by Burt \cite{burt}), but
computational simulation studies of organizational adaptive networks
have begun only recently, e.g., the work by Buskens and Van de Rijt on
the simulation of social network evolution by actors striving for
structural holes \cite{buskens}.

More recent computational models of organizational adaptive networks
are 
hybrids of dynamical networks and agent-based models, where mechanisms
of the coevolution of network topologies and node states can be a lot
more complex and detailed than other more abstract mathematical
models. Such models are therefore hard to study analytically, yet
systematic computational simulations provide equally powerful tools of
investigation. Adaptive network models are still quite novel in
management and organizational sciences, and thus the relevant
literature has just begun to develop.

For example, Dionne et al. \cite{lit32} developed an agent-based model
of team development dynamics, where agents (nodes) exchange their
knowledge through social ties and then update their self-confidence
and trust to other team members dynamically. In their model, the
self-confidence (node state) and trust (link weight) were represented
not by a simple scalar number, but by a complex function defined over
a continuous knowledge domain. Computational simulations illustrated
the nontrivial effects of team network topology and other parameters
on the overall team performance after the team development process.

Another computational model addressing organizational dynamics at a
larger scale was developed by Lin and Desouza \cite{lit34} on the
coevolution of informal organizational network and individual
behavior. In their model, a node state includes behavioral patterns
and knowledge an individual has, and the knowledge is transferred
through informal social links that are changed
adaptively. Computational simulations showed that knowledgeable
individuals do not necessarily gain many connections in the network,
and that when high knowledge diversity exists in the organization, the
network tends to evolve into one with small characteristic path
lengths.

Our most recent work on cultural integration in corporate merger
\cite{lit35} also models organizational dynamics as adaptive networks,
which will be discussed in more detail in Section \ref{culture}.

~

Note that the literature introduced in this section is not meant to be
a comprehensive review of adaptive network research. More extensive
information about the literature and other resources can be found
online \cite{lit36}.

\section{Generative Network Automata}
\label{gna}

In this and the following sections, we present some of our recent work on
computational modeling of adaptive networks and its applications to
complex systems.

To provide a useful modeling framework for adaptive network dynamics,
we have proposed to use graph rewriting systems \cite{GNA1,GNA2} as a
means of uniform representation of state-topology coevolution. This
framework, called Generative Network Automata (GNA), is among the
first to systematically integrate graph rewritings in the
representation and computation of complex network dynamics that
involve both state transition and topological transformation. 

\subsection{Definitions}

A working definition of GNA is a network made of dynamical nodes and
directed links between them. Undirected links can also be represented
by a pair of directed links symmetrically placed between nodes. Each
node takes one of the (finitely or infinitely many) possible states
defined by a node state set $S$. The links describe referential
relationships between the nodes, specifying how the nodes affect each
other in state transition and topological transformation. Each link
may also take one of the possible states in a link state set $S'$. A
configuration of GNA at a specific time $t$ is a combination of states
and topologies of the network, which is formally given by the
following:
\begin{itemize}
\item $V_t$: A finite set of nodes of the network at time $t$. While
  usually assumed as time-invariant in conventional dynamical systems
  theory, this set can dynamically change in the GNA framework due to
  additions and removals of nodes.
\item $C_t:V_t \to S$: A map from the node set to the node state set
  $S$. This describes the global state assignment on the network at
  time $t$. If local states are scalar numbers, this can be
  represented as a simple vector with its size potentially varying
  over time.
\item $L_t:V_t \to \{V_t \times S'\}^*$: A map from the node set to a
  list of destinations of outgoing links and the states of these
  links, where $S'$ is a link state set. This represents the global
  topology of the network at time $t$, which is also potentially
  varying over time.
\end{itemize}

States and topologies of GNA are updated through repetitive graph
rewriting events, each of which consists of the following three steps:
\begin{enumerate}
\item Extraction of part of the GNA (subGNA) that will be subject to
change.
\item Production of a new subGNA that will replace the subGNA selected
above.
\item Embedding of the new subGNA into the rest of the whole GNA.
\end{enumerate}
The temporal dynamics of GNA can therefore be formally defined by the
following triplet $\langle E,R,I \rangle$:
\begin{itemize}
\item $E$: An extraction mechanism that determines which part of the
  GNA is selected for the updating. It is defined as a function that
  takes the whole GNA configuration and returns a specific subGNA in
  it to be replaced. It may be deterministic or stochastic.
\item $R$: A replacement mechanism that produces a new subGNA from the
  subGNA selected by $E$ and also specifies the correspondence of
  nodes between the old and new subGNAs. It is defined as a function
  that takes a subGNA configuration and returns a pair of a new subGNA
  configuration and a mapping between nodes in the old subGNA and
  nodes in the new subGNA. It may be deterministic or stochastic.
\item $I$: An initial configuration of GNA.
\end{itemize}

\begin{figure}
\centering
\includegraphics[height=0.31\textheight]{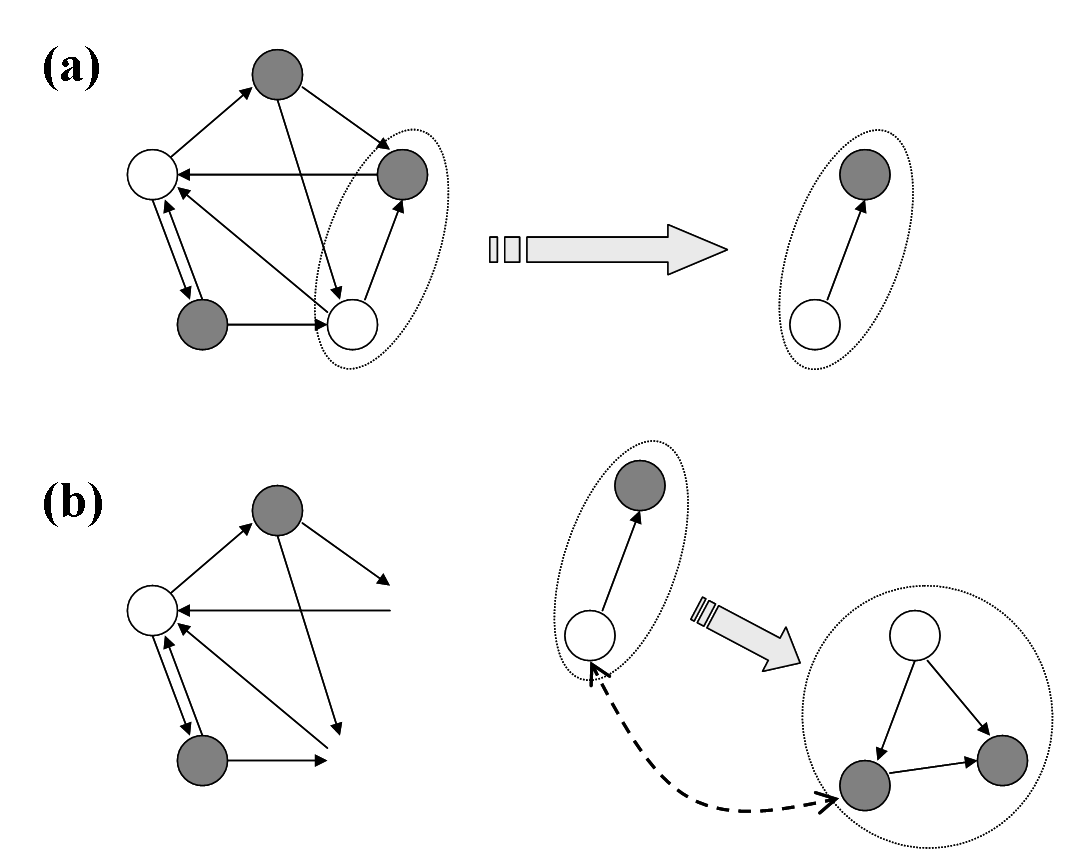}~
\includegraphics[height=0.29\textheight]{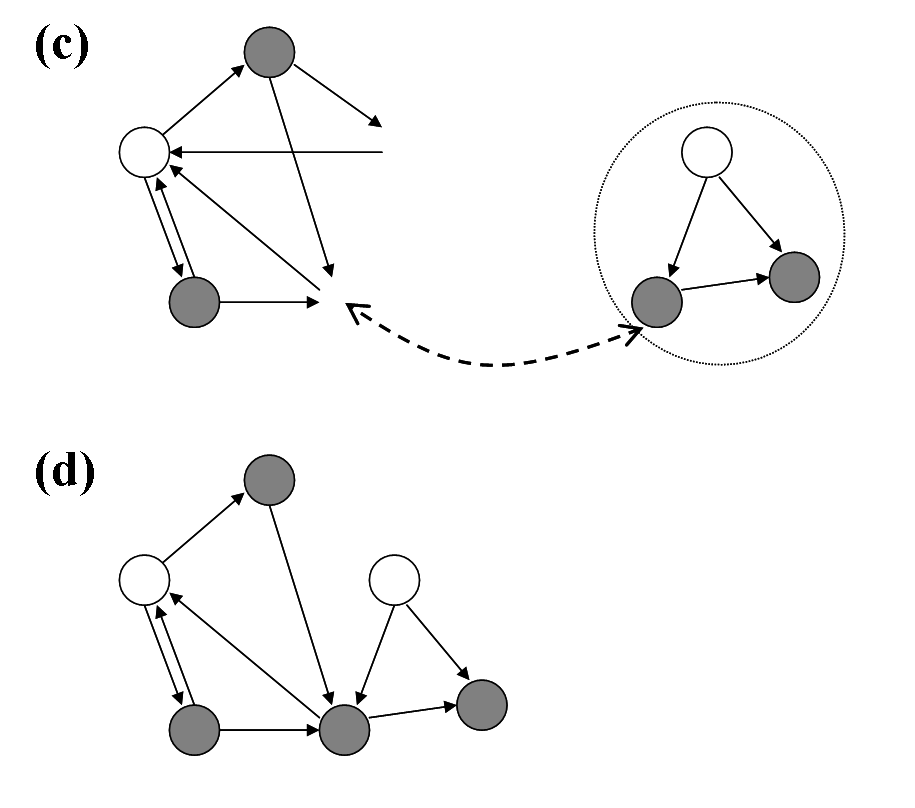}
\caption{GNA rewriting process. (a) The extraction mechanism $E$
selects part of the GNA. (b) The replacement mechanism $R$ produces a
new subGNA as a replacement of the old subGNA and also specifies the
correspondence of nodes between old and new subGNAs (dashed
line). This process may involve both state transition of nodes and
transformation of local topologies. The ``bridge'' links that used to
exist between the old subGNA and the rest of the GNA remain
unconnected and open. (c) The new subGNA produced by $R$ is embedded
into the rest of the GNA according to the node correspondence also
specified by $R$. In this particular example, the top gray node in the
old subGNA has no corresponding node in the new subGNA, so the bridge
links that were connected to that node will be removed. (d) The
updated configuration after this rewriting event.}
\label{rewriting}
\end{figure}

The above $E, R, I$ are sufficient to uniquely define specific GNA
models. The entire picture of a rewriting event is illustrated in
Figure \ref{rewriting}, which visually shows how these mechanisms work
together.

This rewriting process, in general, may not be applied synchronously
to all nodes or subGNAs in a network, because simultaneous
modifications of local network topologies at more than one places may
cause conflicting results that are inconsistent with each other. This
limitation will not apply, though, when there is no possibility of
topological conflicts, e.g., when the rewriting rules are all
context-free, or when GNA is used to simulate conventional dynamical
networks that involve only local state changes but no topological
changes.

\subsection{Uniqueness and Generality of GNA}

The function of the extraction and replacement mechanisms ($E$ and
$R$) may be defined as either deterministic or stochastic, as opposed
to typical deterministic graph grammatical systems
\cite{rozenberg97}. A stochastic representation of GNA dynamics will
be particularly useful when applied to the modeling of real-world
complex network data, in which a considerable amount of random
fluctuations and observation errors are inevitable.

Also, the GNA framework is unique in that the mechanism of subGNA
extraction is explicitly described in the formalism as an algorithm
$E$, not implicitly assumed outside the replacement rules like what
other graph rewriting systems typically adopt (e.g.,
\cite{kurth05}). Such algorithmic specification allows more
flexibility in representing diverse network evolution and less
computational complexity in implementing their simulations,
significantly broadening the areas of application. For example, the
preferential attachment mechanism widely used in network science
to construct scale-free networks is hard to describe with pure graph
grammars but can be easily written in algorithmic form in GNA.

\begin{figure}
\centering \includegraphics[height=0.6\textheight]{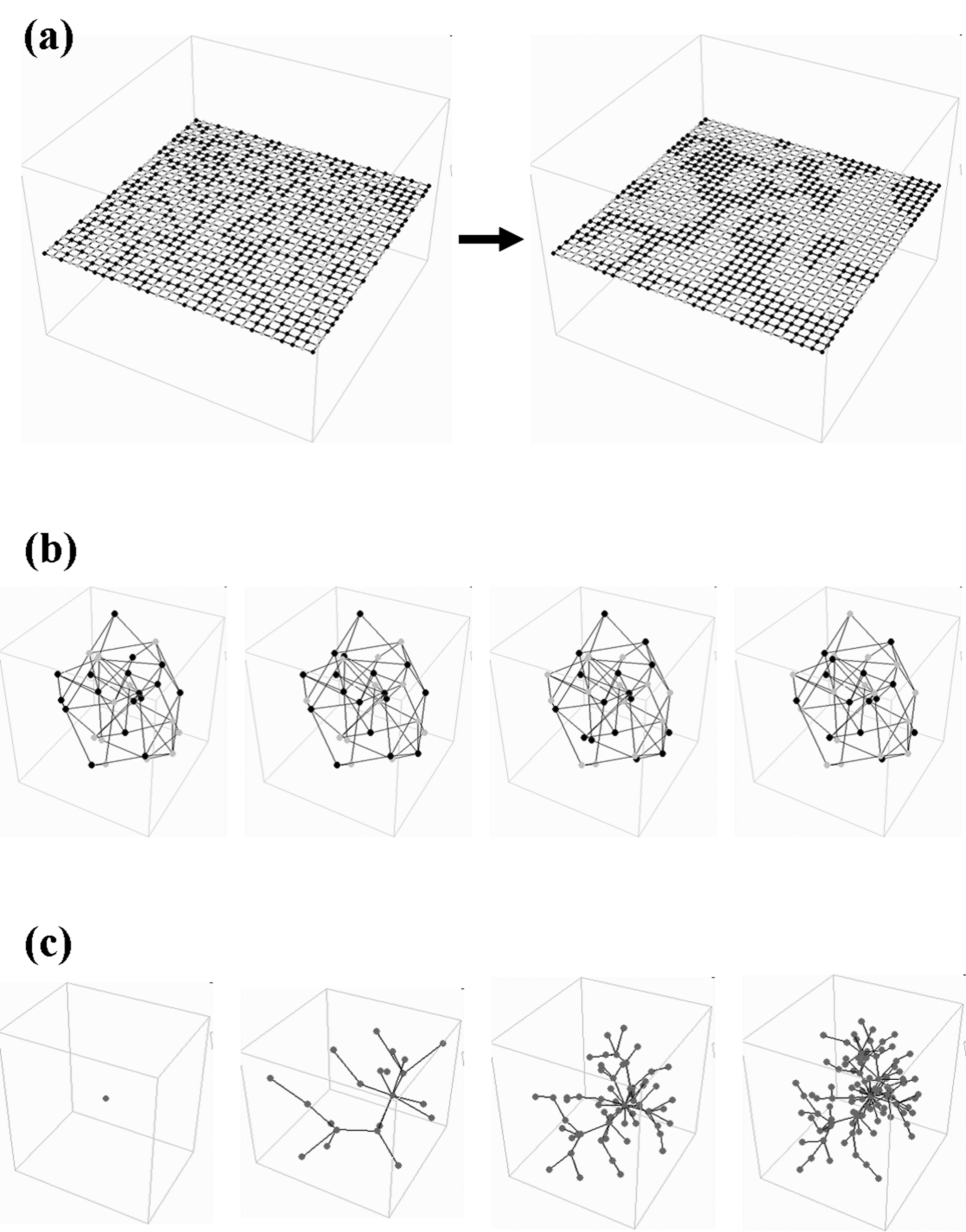}
\caption{Various dynamical network models simulated using GNA. These
  examples were represented in the same format of $\langle E, R, I
  \rangle$ (see text) and simulated using the same simulator package
  implemented in Mathematica. (a) Simulation of asynchronous 2-D
  binary cellular automata with von Neumann neighborhoods and local
  majority rules. Space size: $100 \times 100$. (b) Simulation of an
  asynchronous random Boolean network with $N=30$ and $K=2$. Time
  flows from left to right. Nodes of random Boolean networks are
  non-homogeneous, i.e., they obey different state-transition
  rules. Here each node's own state-transition rule is embedded as
  part of its state, and the replacement mechanism $R$ refers to that
  information when calculating the next state of a node. (c)
  Simulation of a network growth model with the Barab\'{a}si-Albert
  preferential attachment scheme \cite{lit10}. Time flows from left to
  right. Each new node is attached to the network with one link. The
  extraction mechanism $E$ is implemented so that it determines the
  place of attachment preferentially based on the node degrees, which
  causes the formation of a scale-free network in the long run.}
\label{applications}
\end{figure}

The GNA framework is highly general and flexible so that many existing
dynamical network models can be represented and simulated within this
framework. For example, if $R$ always conserves local network
topologies and modifies states of nodes only, then the resulting GNA
is a conventional dynamical network model, including cellular
automata, artificial neural networks, and random Boolean networks
(Figure \ref{applications} (a), (b)). A straightforward application of
GNA typically comes with asynchronous updating schemes, as introduced
above. Since asynchronous automata networks can emulate any
synchronous automata networks \cite{nehaniv04}, the GNA framework
covers the whole class of dynamics that can be produced by
conventional dynamical network models. Moreover, as mentioned earlier,
synchronous updating schemes could also be implemented in GNA for this
particular class of models because they involve only state changes on
each localized node but no topological transformation. On the other
hand, many network growth models developed in network science
can also be represented as GNA if appropriate assumptions are
implemented in the subGNA extraction mechanism $E$ and if the
replacement mechanism $R$ causes no change in local states of nodes
(Figure \ref{applications} (c)).

We also conducted extensive computational experiments of simple
binary-state GNA, which revealed several distinct types of the GNA
dynamics, illustrating the richness and subtleness in the dynamics of
this modeling framework \cite{GNA2}.

\section{Application I: Dynamics of Operational Networks}
\label{Opnet}

In this section, we consider an application of adaptive network models
to socio-technical systems -- a special type of complex systems that
comprise social and technological components or a combination of both
types in one entity \cite{PestovSIAM,PestovVerga}. The services
provided by socio-technical systems can be categorized into two main
functions: (1) detection of a significant event, and (2) execution of
an appropriate response action. The second function involves a
creation of a new network between system components that will be
called upon to execute a response. This new network that dynamically
develops on the nodes of the existing network is termed as an
operational network \cite{PestovCISDA2011}.

In what follows, an adaptive network-based model of the operational
network will be illustrated on an example of the Canadian Arctic
Search and Rescue (SAR) system. The case log of a real incident in the
Arctic will be used to develop and analyze a sample operational
network.

\subsection{SARnet -- an Adaptive Network Model of the Canadian Arctic SAR System}
\label{SARnet} 

The Canadian Arctic Search and Rescue (SAR) system comprises a large
number of highly specialized SAR assets that are trained or designed
to provide a comprehensive range of SAR services. These are Canadian
Coast Guard (CCG) officers, teams of SAR technicians, Joint Rescue
Coordination Centres (JRCCs), aircraft and ships equipped with the
crew, and various information and communication systems. A detailed
description of the Canadian Arctic SAR system is given in
\cite{SARTM}. SARnet is a network model of the Canadian Arctic SAR
system that comprises multiple networks with embedded heterogeneous
agents, where agents are SAR assets. Unlike agents of other typical
social network models, heterogeneous agents of the SAR system cannot
easily be re-trained and replace other agents. The agent specialization
results in a distinctive pattern of network dynamics, as we elaborate
below.

SARnet distinguishes between five classes of agents according to their
specialization (sensor, router, actor, database, and controller; see
Table \ref{tab:agentclass}), six environmental realms in which these
agents operate (maritime, land, air, space, cyber, and cognitive), and
four SAR operational domains according to traditional subdivision of
SAR services (Air, Maritime, Ground, and Joint SAR). In addition,
SARnet represents such agent properties as skill sets, access to
resources, home organizations, and technical specifications.

\begin{table}
\centering
\caption{Summary of Canadian Arctic SAR agent classes.}
\begin{tabular}{|l|l|}
\hline
\textbf{Agent Class}&\textbf{Functions} \\
\hline
Sensor & senses, detects, and passes gathered information \\
Router & distributes the flow of information and enables communication links \\
Actor &  executes a response action as tasked \\
Database & stores and provides access to information \\
Controller & coordinates a response and tasks other agents \\
\hline
\end{tabular}
\label{tab:agentclass}
\end{table} 

The SARnet agent is represented by a string of data of dimension $N$, i.e.,
\begin{equation}
\sigma = [\sigma_1, \sigma_2, \ldots , \sigma_N],  \label{eq:agentrep1}
\end{equation}
where $\sigma_i$ is a binary, categorical or continuous variable that
represents a property of the agent.

We say that agents belong to the same {\em heterotype} if they are
identical in the first several key positions of string $\sigma$. The
distribution of the numbers of agent heterotypes can be used to
measure the agent heterogeneity. If $K$ is the number of heterotypes
and $X_k$ is a fraction of agents of heterotype $k$ $(k =1,\ldots,K)$,
then the network entropy can be defined as follows:
\begin{equation} 
S = -\frac{1}{\ln{K}}\sum_{k=1}^K X_k\ln{X_k}.  \label{eq:entropy}
\end{equation}
In Eq. \ref{eq:entropy}, the network entropy $S$ is normalized by its
maximum value $S_{max}=\ln{K}.$ As follows from Eq. \ref{eq:entropy},
$S\in[0,1]$. The minimum value $S=0$ corresponds to a network composed
of one heterotype. The maximum value $S=1$ corresponds to a network
composed of agents evenly distributed between all $K$ heterotypes
(i.e. $X_k=1/K$ for $k=1,\ldots,K$). As the network entropy approaches
$1$, the agent distribution between heterotypes becomes uniform.

On a day-to-day basis, SAR assets are connected in a standby network,
which represents the standby posture of the system \cite{SARTM}. The
operational network dynamically develops on the nodes of the standby
network in response to a particular SAR incident. It links SAR assets,
which are called upon to provide specified SAR services. A responsible
JRCC initiates a SAR response by appointing one of the controller
agents, as the Search Master. The Search Master is responsible for the
SAR operation in question until closure of the case. The sequence of
services, which will be provided after a distress alert is received,
follows prescribed protocols and procedures, which serve as a
blueprint for tasking SAR assets based on their specialization and
availability. The nature and size of the incident (e.g., location of
the crash site and number of people on board) also determine the
choice of SAR assets being called upon. The dynamics of the
operational network differs from that of the standby network, as the
architecture of the former evolves at the time scale of minutes or
hours instead of months or even years, as in the latter case.

We developed the SARnet simulation software, called OpNetSim, for
automated generation of operational networks, which is described in
\cite{MSV2012}. OpNetSim has its theoretical basis on GNA
\cite{GNA1,GNA2}. The simulation code was developed in Python, and
NetworkX \cite{NetworkX} was used for network representation and
analysis. The network dynamics are described as a set of possible
rewriting events. A rewriting event is defined as an establishment of
a new link between two agents, possibly involving changes of their
states.


Each possible event is specified by the following eight properties:
\begin{enumerate}
\item Conditions: (Optional) Logical expression(s) that indicate when this event can be executable.
\item Source: Agent from which the new link departs.
\item Destination: Agent to which the new link points.
\item Link type: Type of the event (i.e., interaction between the two agents). The following three types are allowed:
	\begin{itemize}
	\item ``Request'': The source agent requests the destination agent for specific information.
	\item ``Flow'': The source agent sends specific information to the destination agent.
	\item ``Task'': The source agent commands the destination agent to do a particular task.
	\end{itemize}
\item Knowledge required: (Optional) List of internal variables the source agent needs to have in order for the event to occur.
\item Knowledge transferred: (Optional) List of internal variables whose values are requested or shared between the two agents during the ``Request'' or ``Flow'' event.
\item Duration: Amount of time the event takes.
\item Duration variation: Amount of stochastic variation for the duration.
\end{enumerate}

OpNetSim reads the set of possible rewriting events given in the above
format. The algorithm of simulation of this network proceeds in the
following steps:
\begin{enumerate}
\item Select all the events that are currently executable (i.e., all conditions are met and the source agent has all the knowledge required).
\item Make the selected events active and set a duration time (with stochastic variation added according to the duration variation property of the event) to their respective internal time counters.
\item Decrease the time counters of all of the active events by a unit time.
\item If the counter of any of those active events hits zero, establish a new directed link from the source agent to the destination agent in the network. Also, depending on the type of the event, update the internal variables of both agents. Then deactivate the event.
\item Repeat the process above until no more executable events exist.
\end{enumerate}

\begin{figure}
\centering
\includegraphics[width=0.5\textwidth]{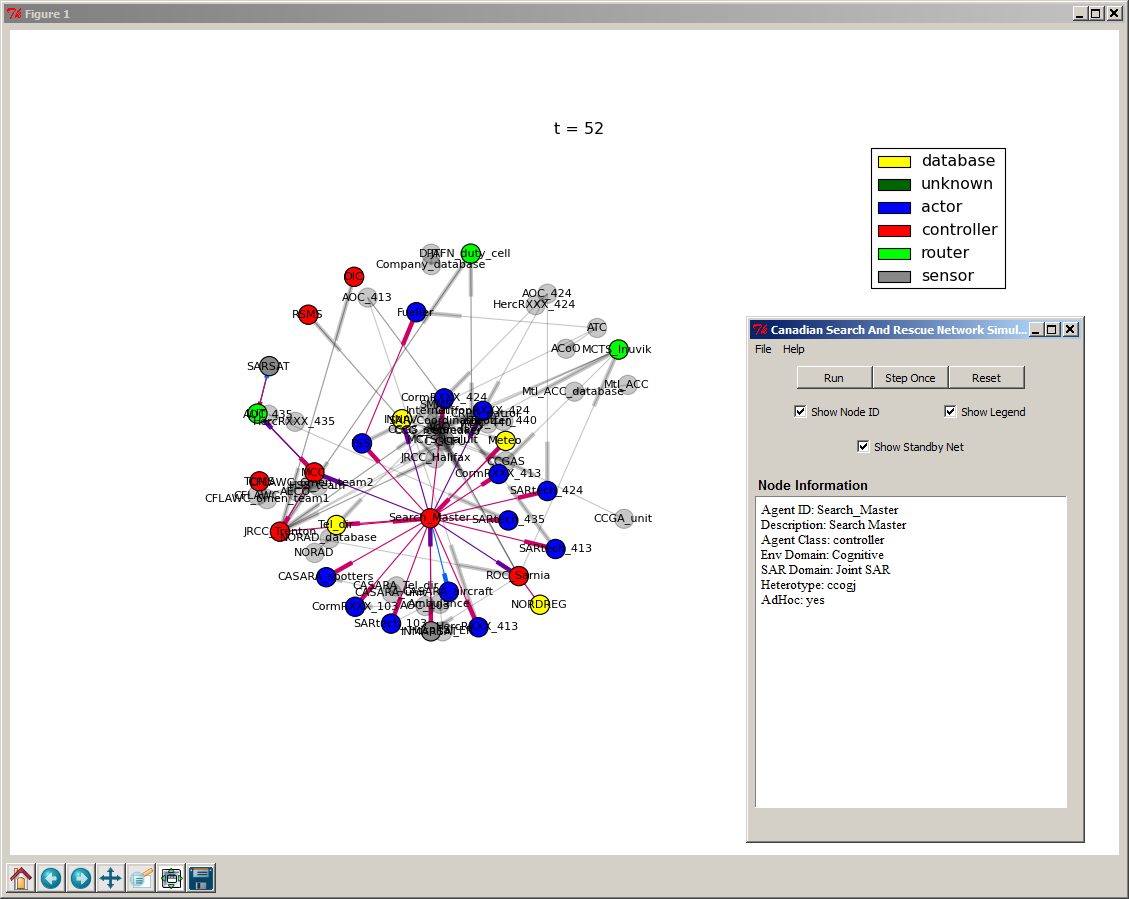}
\caption{Graphical user interface of OpNetSim.}
\label{fig:GUI}
\end{figure}

The operational network will emerge as the simulation progresses and
more agents are connected by information exchange and task
allocation. OpNetSim implements interactive graphical user interface
(GUI) by which the user can operate and inspect the simulation status
(Figure \ref{fig:GUI}).

\subsection{The December 2008 SAR Incident in the Arctic}
\label{A1opnet}

OpNetSim was used to simulate the operational network of a real SAR
incident in the Arctic that occurred in December 2008.

On 7 December 2008 a small two-engine Cessna plane with two people on
board crash-landed in the Arctic approximately 120 nautical miles (nm)
from Iqaluit, Nunavut. Three mayday calls were intercepted by a
commercial aircraft and by a Royal Canadian Air Force (RCAF) aircraft,
and then relayed to JRCC Halifax. (JRCC Halifax, located in Halifax
NS, is a JRCC responsible for that sector of the Arctic.) The Canadian
SAR system mounted a response to the incident, which involved three
RCAF SAR squadrons, Canadian Coast Guard (CCG) resources (including
database resources and marine communication systems), regional units
of the Royal Canadian Mounted Police (RCMP) and Civil Air Search and
Rescue Association (CASARA), local police, air-ground-air
communication systems, and private-sector assets. One of the CCG
officers on duty was appointed as the Search Master to coordinate the
SAR operation. In less than 18 hours from the time of the first mayday
call, the two survivors were rescued (with mild frostbites, otherwise
in good condition).

The concept of the operational network was used to represent and
analyze the operational architecture of the SAR response to this
incident, and to identify factors contributing to a successful
outcome.

The agent heterogeneity was identified as the main driving mechanism
for the development of the operational network. Such agent attributes
as agent class, realm, and SAR domain influenced the formation of
network architecture. The agents' skill sets and access to resources
as well as the crash-site information also plays a role in shaping the
operational network. The architecture of the resulting network evolved
at the time scale of minutes or hours instead of months or even years,
as in the case of the standby network.

\begin{figure}[t]
  \centering
  \includegraphics[width=0.8\textwidth]{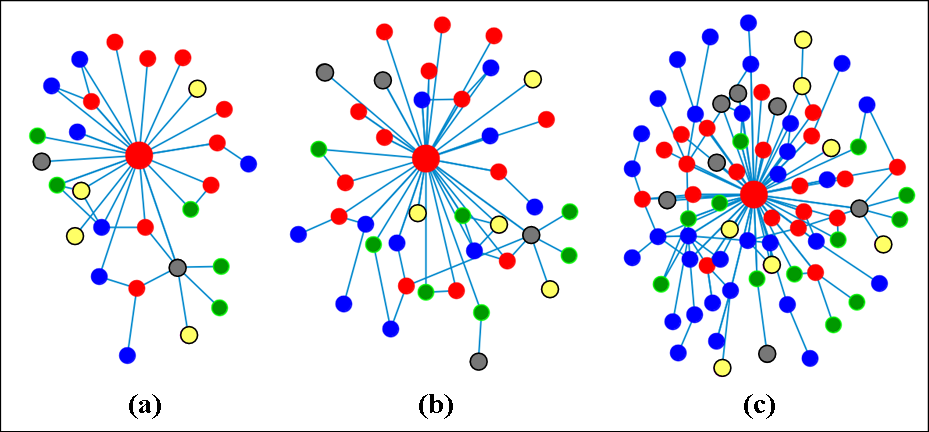}
  \caption{The development of the operational network drawn based on
    the real SAR incident in the Arctic in December 2008: (a) 1 hour,
    (b) 3 hours, and (c) 18 hours after the response
    initiation. Controllers are shown by red, actors by blue, sensors
    by gray, routers by green, and databases by yellow circles. The
    Search Master is shown by an enlarged red circle. The ORA network
    analysis and visualization software \cite{ORAguide} was used to
    visualize the networks.}
  \label{fig:A1Opnetx3}
\end{figure}

Figure \ref{fig:A1Opnetx3} shows the snapshots of the actual
operational network one, three and 18 hours after the response
initiation, respectively.

The number of agent heterotypes was increasing in the course of the
SAR response, meaning that the network heterogeneity was also
increasing. At the same time, the distribution of agents between
heterotypes became less balanced, as follows from a decline in
normalized entropy after 6 hours of the response. Table
\ref{tab:dynamics} summarizes the development of the operational
network after 1, 3, and 18 hours.

\begin{table}[ht]
\centering
\caption[Network development]{Network development after 1, 3, and 18 hours.}
\begin{tabular}{|l|l|l|l|}
\hline
 & {\bf\small 1 hour} & {\bf\small 3 hours} & {\bf\small 18 hours}\\
\hline
{\small Node count:} & 28 & 40 & 79 \\
\hline
{\small Link count:} & 41 & 64 & 140 \\
\hline
{\small Link weight:} &  &  &  \\
{\small $\,\,\,\,$ Max.} & 4 & 6 & 12 \\
{\small $\,\,\,\,$ Min.} & 1 & 1 & 1 \\
{\small $\,\,\,\,$ Average} & 1.37 & 1.48 & 1.96 \\
\hline
{\small Network composition:} &  &  & \\
{\small $\,\,\,\,$ Actors:} & 7 (25\%) & 10 (25\%) & 32 (40.5\%) \\
{\small $\,\,\,\,$ Controllers:} & 10 (36\%) & 15 (37.5\%) &	21 (26.5\%) \\
{\small $\,\,\,\,$ Databases:} & 4 (14\%) & 4 (10\%) & 7 (9\%) \\
{\small $\,\,\,\,$ Routers:} & 5 (18\%) & 7 (17.5\%) &	12 (15\%) \\
{\small $\,\,\,\,$ Sensors:} & 2 (7\%) & 4 (10\%) &	7 (9\%) \\
\hline
{\small Number of heterotypes:} & 14 & 17 &	23 \\
\hline
{\small Network entropy (normalized):} & 0.92520 & 0.89210 &	0.86274 \\
\hline
\end{tabular}
\label{tab:dynamics}
\end{table}

According to our analysis, all major players were quickly identified
and added to the network at early stages of its development. By the
end of the first hour of the response, the operational network
included 28 agents and 41 links, i.e. 30\% of the final network. After
the first three hours, 40 agents and 64 links, i.e. nearly 50\% of the
final net, were in place. By the time when the survivors were rescued
(i.e. 18 hours after the response initiation), more than 80\% of the
operational net had developed.

The Search Master (which was an isolate in the standby network)
quickly became the most influential entity of the operational network,
coming first in all node-level measures, including standard Social
Network Analysis measures of degree centrality and extended measures
of cognitive demand and shared situation awareness (see
\cite{ORAguide} for measure definitions).

After 18 hours, the Search Master's sphere of influence\footnote[5]{A
  Sphere of influence of a node is a sub-network of radius 1 that
  includes all nodes to whom that node has direct connections plus
  connections between those nodes.} encompassed 67\% of the entire
operational network. For comparison, the sphere of influence of the
next most influential node contained about 14\% of the network. The
Search Master had direct interactions with 44 out of 78 other
agents. (In total, 79 agents were included in the operational
network.) This value was almost an order of magnitude higher than that
of the second-ranked agent. Average communication speed between any
two nodes within the Search Master's sphere of influence (or 67\% of
the network) was close to $0.5$, and the average speed with which the
Search Master interacted with 67\% of all agents was $1.0$ -- the
maximum value for this measure.

High centralization of the operational net was identified as a
contributing factor to the operational effectiveness. However, it can
also be viewed as a vulnerability factor. According to our analysis
results, the removal of the Search Master will lead to maximum network
fragmentation when almost 80\% of SAR assets will become
disconnected. Moreover, there was no other entity in the network
capable of assuming the leadership role in the SAR response in
question. A detailed summary of network analysis results can be found
in \cite{SARTM}.

\begin{figure}[t]
  \centering
  \fbox{
   ~~
  \includegraphics[width=0.24\textwidth]{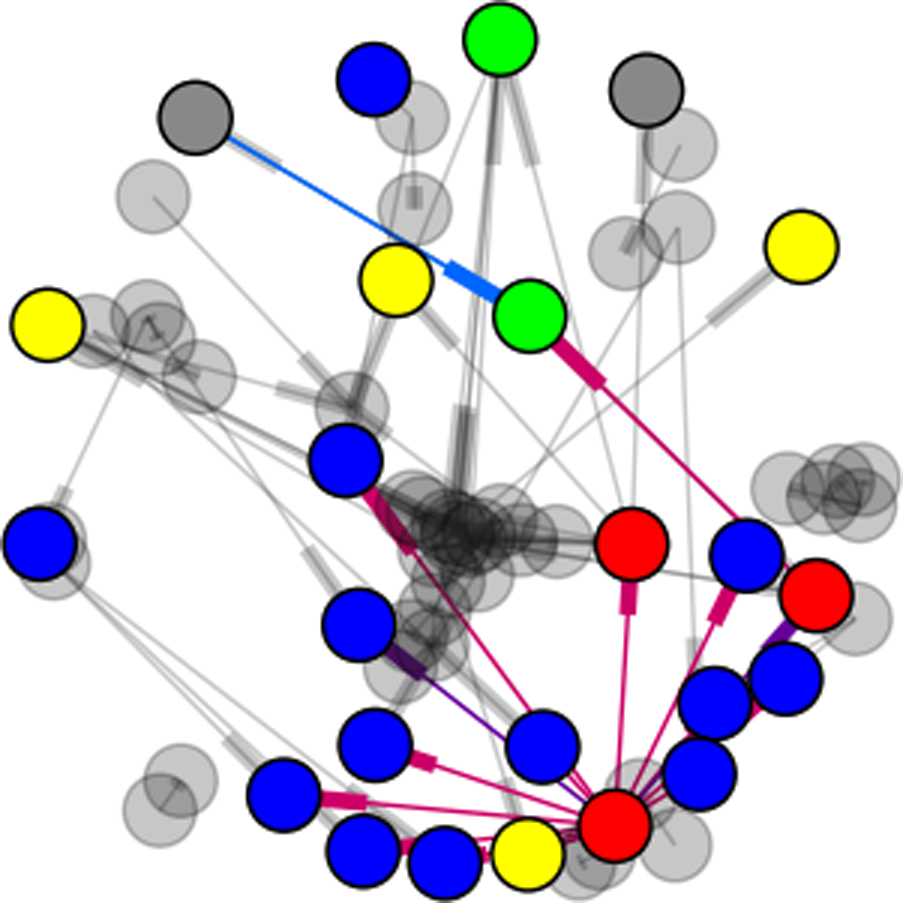}~~~
  \includegraphics[width=0.24\textwidth]{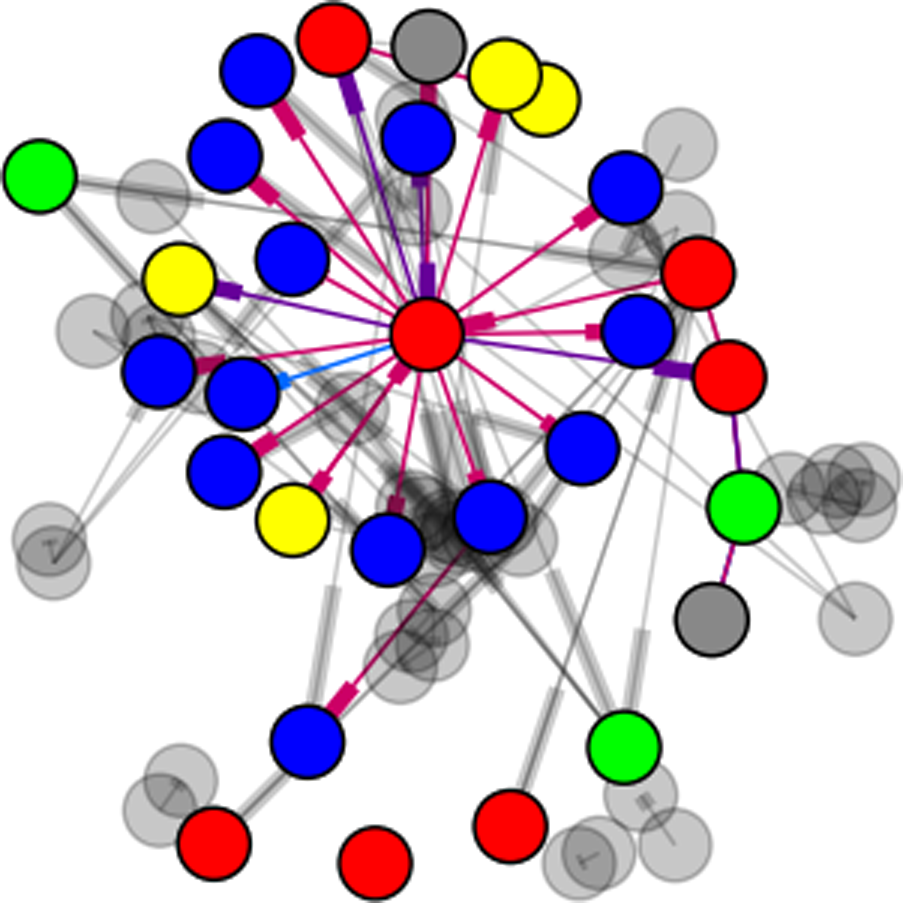}~~~
  \includegraphics[width=0.24\textwidth]{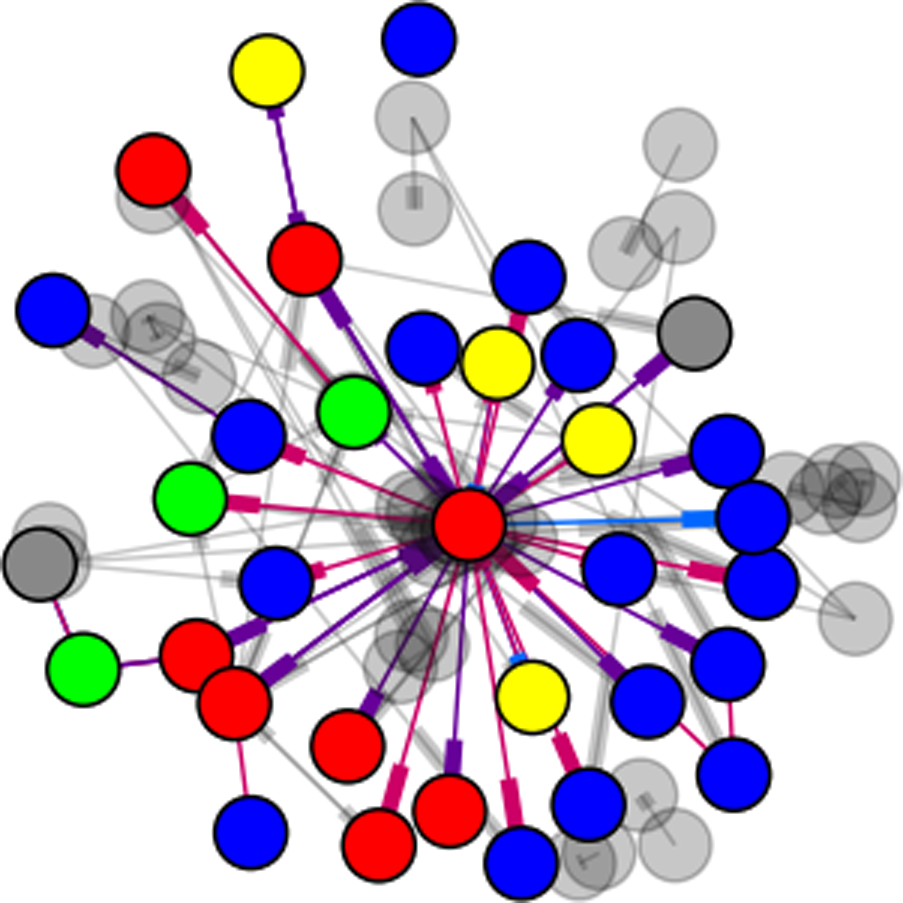}
  ~~
  }
  \caption{The development of the operational network simulated using
    OpNetSim. Time flows from left to right. The gray nodes and links
    represent the standby network, while colored nodes are the
    activated ones that form the dynamically changing operational
    network. Visualization is done using Python with NetworkX. Node
    color schemes are the same as in Figure \ref{fig:A1Opnetx3}.}
  \label{fig:OpNetSimResults}
\end{figure}

We examined the actual log of inter-agent communications during this
SAR incident, and manually reconstructed the rewriting rules that
drove the operational network formation. OpNetSim was then used to
simulate the temporal development of the operational network under
several hypothetical scenarios. Figure \ref{fig:OpNetSimResults} shows
snapshots of the simulated operational network produced by
OpNetSim. Since the simulation algorithm involves stochasticity, the
topology of the simulated network does not exactly match the actual
one, but the general trend of increasing agent heterogeneity and
concentration on the Search Master node were correctly represented in
this model.

\section{Application II: Automated Rule Discovery from Empirical Network Evolution Data}
\label{modeldiscovery}

In the previous section, we developed an adaptive network model based
on our knowledge and understanding about local dynamics of node and
link interactions. In the meantime, it has remained an open question
how one could derive dynamical rules of an adaptive network model
directly from a given empirical data of network evolution.

\begin{figure}
\centering
\includegraphics[height=0.8\textheight]{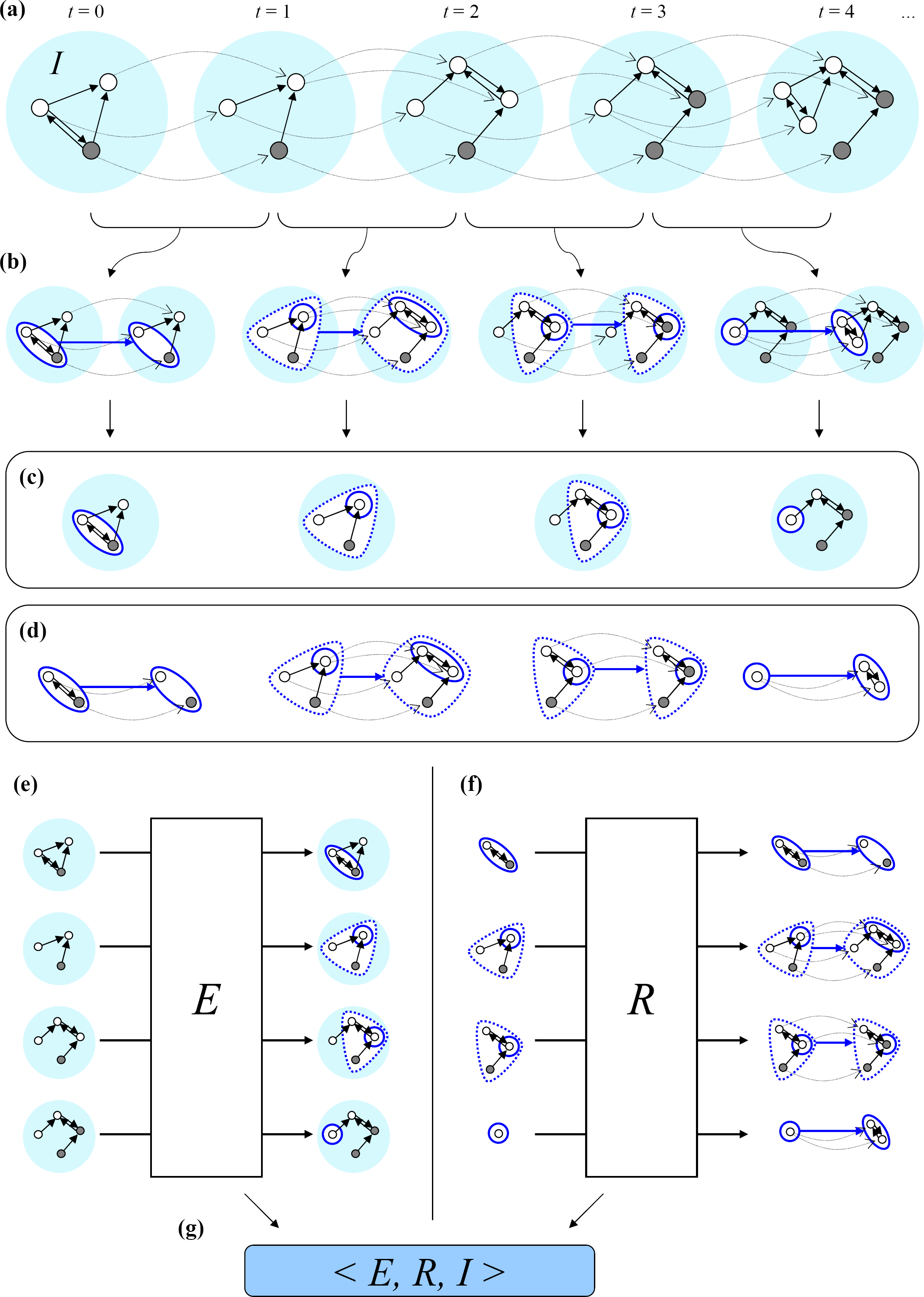}
\caption{Overview of the proposed algorithm for automatic discovery of
  GNA rewriting rules. (a) Original network evolution data starting
  with the initial configuration $I$. (b) Detection of rewriting
  events at every time step. (c) Training data for the extraction
  mechanism $E$. (d) Training data for the replacement mechanism
  $R$. (e, f) Construction of models of $E$ and $R$ based on the
  training data. (g) Final GNA model.}
\label{algorithm}
\end{figure}

In this section, we describe an algorithm that automatically discovers
a set of dynamical rules that best captures state transition and
topological transformation expressed in the empirical data
\cite{BIONETICS2010}. Network evolution is formulated using the GNA
framework and the subnetwork extraction and replacement phases are
analyzed separately. Within the scope of this paper, we will simplify
the problem by requiring the data to satisfy the following:

\begin{enumerate}
\item A given data set is a series of configurations of labeled
  directed or undirected networks in which labels (states) and
  topologies coevolve over discrete time steps (Figure
  \ref{algorithm}(a)).
\item The data set contains information about the correspondence of
  nodes between every pair of two successive time points (Figure
  \ref{algorithm}(a)).
\item States are discrete, finite, and assigned only to nodes, not to
  links.
\item Changes that take place between successive time points are
  reasonably small so that they can be identified as one small network
  rewriting event per each time step.
\item The extraction mechanism $E$ and the replacement mechanism $R$
  are memoryless, i.e., they produce outputs solely based on inputs
  given to them.
\end{enumerate}

We note that the GNA framework has a significant advantage for the
algorithm design. It formulates the network evolution using two
separate phases, i.e., the extraction of subGNA (performed by $E$) and
its replacement (performed by $R$). Therefore, the estimation and
construction of models of $E$ and $R$ can be conducted independently
and concurrently using separate training data sets, which will make
the algorithm simple and tractable.

\subsection{Proposed Algorithm}

A general procedure of the proposed algorithm is as follows (Figure
\ref{algorithm}):

\begin{enumerate}
\item Preprocess the original network evolution data using
  data-dependent heuristics, if necessary, so that they meet all the
  aforementioned requirements.

\item Detect the difference between each pair of configurations at two
  successive time points $(G_t, G_{t+1})$ and represent it as a
  rewriting event $s_t \Rightarrow r_t$ (Figure \ref{algorithm}(b)),
  where $s_t$ is a subGNA to be replaced, $r_t$ is another subGNA that
  replaces $s_t$, and ``$\Rightarrow$'' denotes correspondence from
  nodes in $s_t$ to nodes in $r_t$.

\quad The difference between two configurations ($G_t = \langle V_t,
C_t, L_t \rangle, G_{t+1} = \langle V_{t+1}, C_{t+1}, L_{t+1}
\rangle$) will be detected in the following way:
\begin{enumerate}
\item Let $A$ be a set of nodes in $G_t$ which disappeared in
  $G_{t+1}$ ($A = \{x | x \in V_t \wedge x \not\in V_{t+1}\}$).

\item Let $B$ be a set of nodes in $G_{t+1}$ which did not exist in
  $G_t$ ($B = \{x | x \in V_{t+1} \wedge x \not\in V_t \}$).

\item Add to $A$ and $B$ all the nodes whose states or neighbors
  changed between $G_t$ and $G_{t+1}$ ($D = \{x | x \in V_t \wedge x
    \in$ $V_{t+1} \wedge (C_t(x) \neq C_{t+1}(x) \vee L_t(x) \neq
    L_{t+1}(x))\}$, $A = A \cup D$, $B = B \cup D$).

\quad At this point, $A$ and $B$ contain the nodes that experienced some
changes (enclosed by solid lines in Figure \ref{algorithm}(b)).

\item Add to $A$ and $B$ all the nodes which have a link to any of the
  nodes in $A$ ($D' = \{x | x \in V_t \wedge x \in V_{t+1} \wedge
    L_t(x) \cap$ $A \neq \{\emptyset\}\}$, $A = A \cup D'$, $B = B \cup
  D'$).
 
\quad The above step includes in $A$ and $B$ additional nodes that may have
influenced the rewriting event (enclosed by dashed lines in
Figure \ref{algorithm}(b)).

\item Let $s_t$ and $r_t$ be subgraphs of $G_t$ and $G_{t+1}$ induced
  by nodes in $A$ and $B$, respectively.
\end{enumerate}

Then the detected rewriting event is represented as $s_t \Rightarrow
r_t$ , where ``$\Rightarrow$'' is the set of all the node
correspondences between $s_t$ and $r_t$ present in the original data.

\item Construct a model of the extraction mechanism $E$ by using ${
  (G_t, s_t) }$ as training data, where $G_t$ is the input given to
  $E$ and $s_t$ the output that $E$ should produce
  (Figure \ref{algorithm}(c),(e)).

\quad This step is the most challenging part in this algorithm development
effort. The task to be achieved in this step is to identify an unknown
mechanism that chooses a subset of a given set of nodes. Exact
identification of an unknown computational mechanism is theoretically
not possible in general.

\quad Here, we will assume several predefined candidate mechanisms
(e.g., random selection, preferential selection based on node degrees,
motif-based selection, etc.)  and calculate the likelihood for each
extraction result given in the training data to occur with each
candidate mechanism. This calculation will be conducted and multiplied
sequentially over the whole training data to evaluate how likely the
given training data could result from each of the candidate
mechanisms. If a mechanism includes parameters, they will be optimized
to attain the maximal probability. Then the mechanism with the highest
likelihood will be returned as the estimated mechanism of $E$.

\item Construct a model of the replacement mechanism $R$ by using ${
  (s_t, s_t \Rightarrow r_t) }$ as training data, where $s_t$ is the
  input given to $R$ and $s_t \Rightarrow r_t$ the output $R$ should
  produce (Figure \ref{algorithm}(d),(f)).

\quad In this step, the task can be achieved in a much simpler manner
than in step (3) (though technically it still remains identification
of an unknown mechanism). This is because a single rewriting event
typically involves just a few nodes so the number of possible inputs
given to the replacement mechanism $R$ is virtually finite in contrast
to the number of possible inputs to $E$ that is virtually
infinite. Therefore we use straightforward pattern matching methods to
construct a model of $R$ from the data. Specifically, the algorithm
will construct $R$ as a simple procedure that searches for a rewriting
event in the training data whose left hand side matches the given
input. If there is only one such event found, the event itself will be
the output of $R$. If multiple events are found, the output will be
determined either deterministically (e.g., event with greatest
frequency) or stochastically (e.g., random selection with weights set
proportional to event frequencies). Or, if no event is found, either
identity (``input $\Rightarrow$ input''; no change) will be returned
or seek similar events will be sought using partial graph matching
schemes.

\item Construct a complete GNA model by combining the results of the
  above steps (3) and (4) together with the initial configuration $I$
  (Figure \ref{algorithm}(g)).
\end{enumerate}

\subsection{Software Implementation}

We have designed the details of the algorithm described above and implemented them in Python with
NetworkX and GraphML. The software, called PyGNA \cite{schmidt11,schmidt12}, is
designed to automatically discover a set of dynamical rules that best
captures both state transition and topological transformation in the
data of spatio-temporal evolution of a complex network. PyGNA is still
at its alpha stage, but is publicly available from
SourceForge.net\footnote{http://sourceforge.net/projects/gnaframework/}.

\begin{figure}[tp]
\centering
\includegraphics[width=\textwidth]{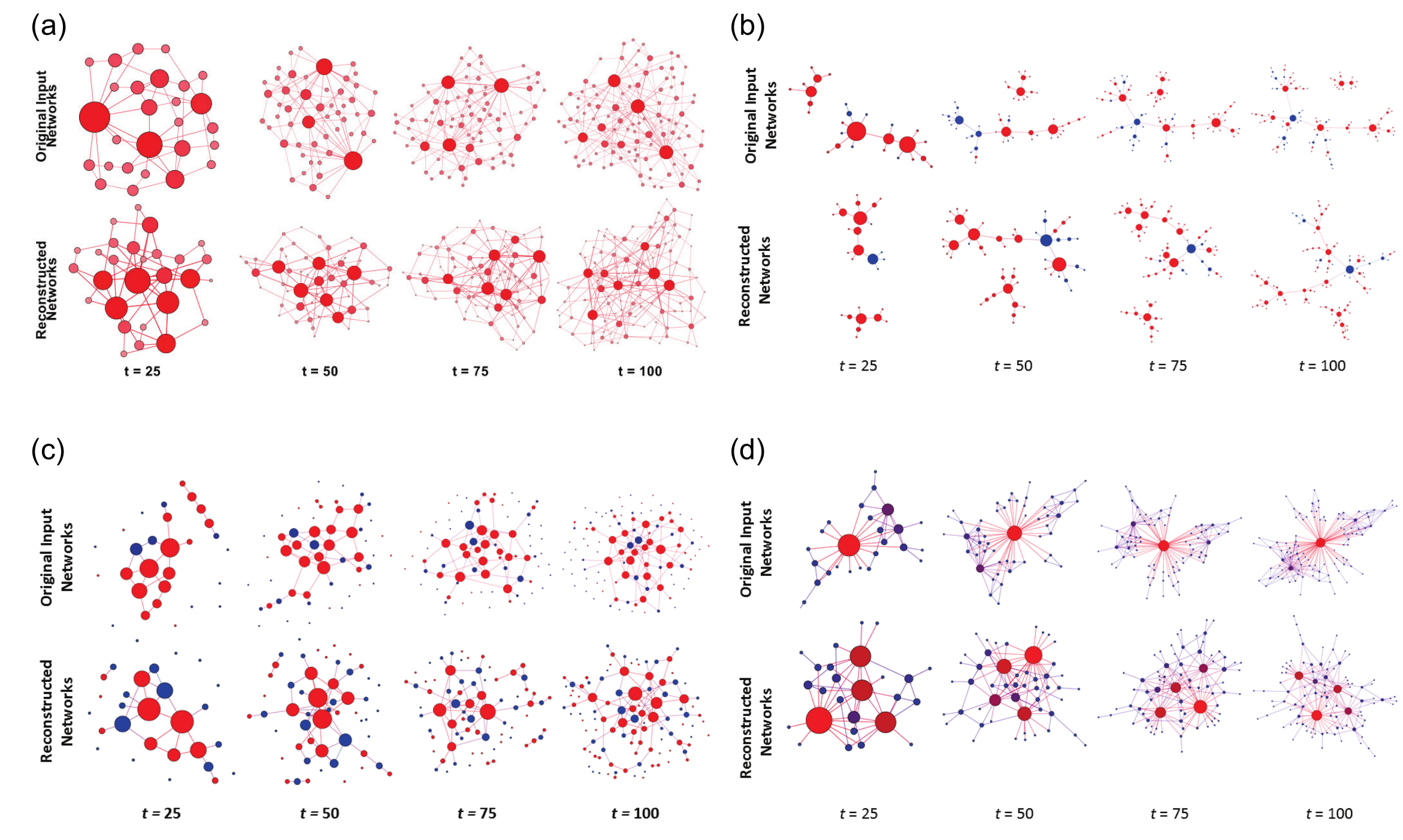} 
\caption{Results of experiments to reconstruct dynamical network
  models from artificially generated network data. Sizes of nodes
  represent their relative degrees. $t$ represents the number of
  iterations. (a) Barabasi-Albert network. (b) Degree-state
  network. (c) State-based network. (d) Forest fire network. See text
  for details.}
\label{PyGNA-results}
\end{figure}

We conducted preliminary experiments applying PyGNA to data generated
by abstract adaptive network models, in order to test if it could
correctly identify the actual network generation mechanisms used to
reproduce the input data. The following four abstract network models
were used as inputs to PyGNA:
\begin{enumerate}[(a)]
\item Barabasi-Albert network, grown using the standard degree-based
  preferential attachment method \cite{lit10}.
\item ``Degree-state'' network, grown by degree-based preferential
  attachment whose mechanism is influenced by the randomly determined
  state of the newcomer node. The state of the target node could also
  be altered by the attachment.
\item ``State-based'' network, grown by repeated random edge addition
  between a node that has a particular state (shown in red in Figure
  \ref{PyGNA-results}) and any other randomly selected node. New
  isolated nodes are also continuously introduced to the network, with
  a randomly selected state.
\item ``Forest fire'' network, generated by the method proposed in
  \cite{ff}.
\end{enumerate}

Figure \ref{PyGNA-results} shows typical results that visually compare
the original input networks and the networks reconstructed by
PyGNA. For the Barabasi-Albert (a), degree-state (b) and state-based
(c) networks, both input and reconstructed networks have visually
similar structures. PyGNA also correctly identified that the growth of
those networks was determined by degrees (for (a)), degrees and states
(for (b)), and states (for (c)). For the forest fire network (d),
however, PyGNA failed to capture the unique topological
characteristics of the original input network, because of the
complexity in the original network generation method.

We also quantified the accuracy of the reconstructed network models by
measuring the distance of probability distributions of extracted
subgraphs between original and simulated networks. Specifically, for
the original input data and the reconstructed network simulation
results, we counted how many times each of the different kinds of
subgraphs was selected for graph rewriting events, and then computed
the Bhattacharyya distance \cite{bd1943} between the two
distributions, defined as
\begin{equation}
D_B = - \ln \sum_{s \in S} \sqrt{p(s) q(s)} ,
\end{equation}
where $s$ is the unique subgraph, $S$ the set of all extracted
subgraphs, and $p(s)$ and $q(s)$ the probability distributions of
subgraphs extracted for rewriting in the input network and in the
reconstructed network, respectively. $D_B = 0$ means the two
distributions were exactly the same, while higher value of $D_B$ means
they are far apart.

\begin{figure}[tbp]
\centering
\includegraphics[width=0.65\textwidth]{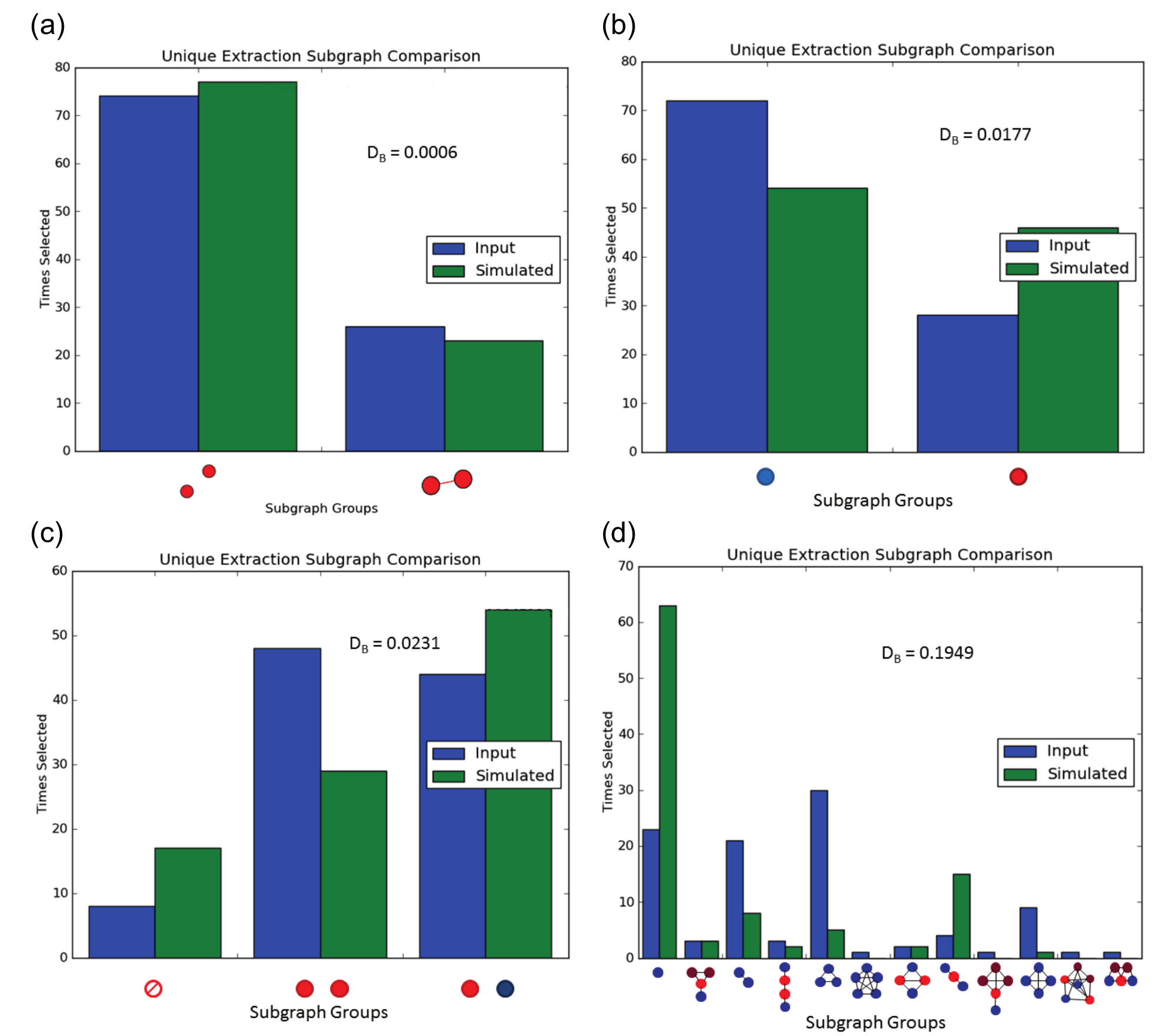} 
\caption{Accuracy of reconstructed network models measured using the
  Bhattacharyya distance ($D_B$) of probability distributions of
  extracted subgraphs between original and simulated
  networks. Horizontal axes represent different kinds of extracted
  subgraphs. (a) Barabasi-Albert network. (b) Degree-state
  network. (c) State-based network. (d) Forest fire network. See text
  for details.}
\label{PyGNA-results2}
\end{figure}

The results are summarized in Figure \ref{PyGNA-results2}. For the
Barabasi-Albert network (a), the low $D_B$ value indicates that the
simulated network is indeed very close to the original input
network. The $D_B$ value was a little higher for the degree-state (b)
and state-based (c) networks, but the overall trends of the extracted
subgraph distributions were generally in agreement between the input
and simulated networks. For the forest fire network, however, the
extraction mechanism selected by PyGNA was over-choosing certain
subgraphs and was unable to generate many subgraphs seen in the input
data, resulting in the apparent topological difference seen in Figure
\ref{PyGNA-results}. The $D_B$ value for this case is therefore
substantially larger than the other three cases.

These preliminary results tell us that the current algorithm in PyGNA
is effective for certain types of networks while still limited for the
analysis of others, especially those that involve pure randomness
and/or mesoscopic topological structures such as motifs. We are
currently revising and expanding our algorithm by addressing these
issues in order to improve the performance of PyGNA.

\section{Application III: Cultural Integration in Corporate Merger}
\label{culture}

The final example we present is a computational model of cultural
integration taking place on a dynamically changing adaptive social
network when two firms merge into one. This example is more complex
than the previous two, firstly because the model has continuous link
weights that adaptively change due to node state dynamics, but more
importantly because the node states are far more complex than in the
previous two models, in order to represent complex sociocultural
aspects of agents. In this sense, this example can be better
understood as a hybrid of agent-based models and adaptive network
models.

It is recognized that cultural integration, or sharing a common
corporate culture, is crucial for the success of corporate
mergers. However, previous studies have been limited to firm-level
analyses only, while cultural adoption and diffusion in a merged firm
actually occurs among individuals. We thus explored, using the
computational model, how cultural integration emerges from the
patterns of dynamic social interactions among individuals
\cite{lit35}. Our computer simulation model is an agent-based
model operating on a dynamic network structure, where individuals
(nodes) exchange elements of a corporate culture with others who are
connected to it through social ties (links). In this model, we set two
merging firms, A and B, each consisting of 50 individuals. Our goal is
to find initial network structures that promote or impede post-merger
cultural integration. Although the number of individuals in the firms
is by far smaller than that of publicly traded firms, we found that
this parameter has a negligible impact on the simulation results when
network density is kept at the same level.

\subsection{Representations of Corporate Cultures}

We represent a corporate culture as a vector in a multi-dimensional
continuous cultural space. The cultural space is composed of several
cultural dimensions; each dimension represents an element of a
corporate culture. We set 10 cultural dimensions for the cultural
space; this number is founded on previous empirical studies of
corporate culture. For example, O'Reilly et al. \cite{oreilly1991},
who investigated eight large U.S. public accounting firms, found eight
dimensions of organizational cultures: innovation, attention to
detail, outcome orientation, aggressiveness, supportiveness, emphasis
on rewards, team orientation, and decisiveness. Likewise, Chatterjee
et al. \cite{chatterjee1992} measured cultural distance perceived by
the top management teams of acquired firms across seven dimensions of
organizational cultures: innovation and action orientation,
risk-taking, lateral integration, top management contact, autonomy and
decision making, performance orientation, and reward
orientation. Therefore, setting 10 dimensions as elements of corporate
cultures would be a more conservative approach.

In our model, we characterize the distance between two cultures by the
Euclidean distance between two vectors in the cultural space. The
average cultural difference between the two merging firms is
characterized as the average cultural distance between two
individuals---one in Firm A and the other in Firm B. If the value of
this measurement is large, the corporate culture that individuals
perceive in Firm A is, on average, far different from that in Firm
B. We initialized the individual cultural vectors as follows: First,
two cultural ``center'' vectors were created for the two merging
firms, and these center vectors were separated by 3.0 (in an arbitrary
unit) in the cultural space.  Then individual cultural vectors were
created for individuals in each firm by adding a small random number
drawn from a normal distribution with a mean of 0 and a standard
deviation of 0.1 (in the same unit used above) to each component of
the cultural center vector of that firm. This setting creates an
initial condition where the average between-firm cultural difference
is approximately seven times larger than the average within-firm
cultural difference.

\subsection{Adaptive Changes of Cultural States and Tie Strengths}

Individuals in our model are connected to each other through directed
social ties. A tie going from one individual to another works as a
conduit that can transmit, from the origin node to the destination
node, information and knowledge that include the elements of their
corporate cultures. Each tie has a weight associated with it, called
tie strength in the social network literature
\cite{granovetter1973,wasserman-faust1994}. The range of possible tie
strength values is bounded between 0 and 1.  Corporate cultures
diffuse among individuals through their ties. The algorithm for
simulating the dynamics of cultural diffusion, and subsequent social
network changes, is as follows.

One iteration in a simulation consists of simulations of individual
actions for all individuals in a sequential order (therefore there are
always 100 individual actions simulated in each iteration). When it is
its turn to take an action, an individual first selects an information
source. For 99\% of the time, the individual chooses the information
source from its local in-neighbors, that is, the nodes from which
directed ties are coming to the individual. The probability for a
neighbor to be selected as the information source is proportional to
the strength of the tie that connects the neighbor to the individual;
this represents that individuals tend to listen more often to others
whom they trust more or with whom they have stronger
connections. Otherwise (with a 1\% chance), the individual chooses as
the information source any individual in the connected component in
which the individual belongs. If there is no existing tie from the
randomly selected source to the individual, a new tie with a very weak
strength (0.01) will be created between them. This represents an
informal, incidental communication, like a ``water-cooler''
conversation within an organization.

\begin{figure}
\centering
\includegraphics[width=0.5\textwidth]{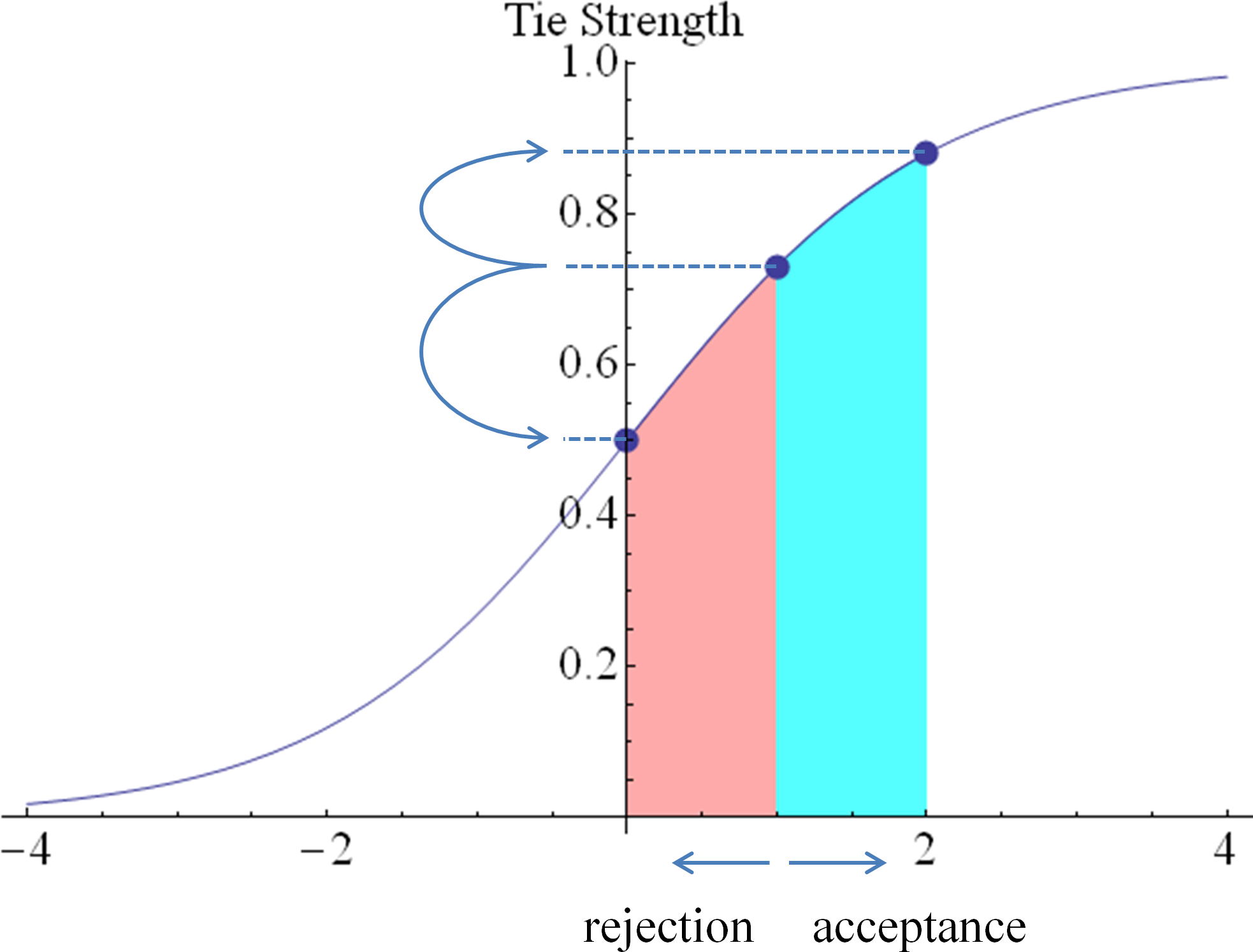}
\caption{Adaptive changes of tie strength caused by cultural
  acceptance or rejection (from \cite{lit35}).}
\label{updating}
\end{figure}

Once the information source is selected, the individual receives the
source's cultural vector and then measures the distance between the
received cultural vector and its own cultural vector. With a
probability that decreases monotonically with increasing cultural
distance, the individual accepts the received culture. The probability
of acceptance, $P_A$, is mathematically represented as
\begin{equation}
P_A(d) = \left(\frac{1}{2}\right)^{d/d_c} ,
\end{equation}
where $d$ is the distance between the two cultural vectors and $d_c$
is the characteristic cultural distance at which $P_A$ becomes
50\%. We used $d_c = 0.5$ for our simulations. If the individual
accepts the received cultural vector, it adopts the mean of the two
vectors (i.e., the sum of the two vectors divided by 2) as its new
cultural vector, and the strength of the tie from the source to the
individual is increased by the following formula:
\begin{equation}
S_{\mathrm{new}} = \mathrm{logistic} (\mathrm{logit}(S_{\mathrm{current}}) + 1 )
\end{equation}
Here $S_{\mathrm{current}}$ and $S_{\mathrm{new}}$ are the current and
updated tie strengths, respectively (this formula guarantees that the
tie strength is always constrained between 0 and 1). On the other
hand, if the individual rejects the received cultural vector, its own
vector will not change, and the tie strength is decreased by the
following formula:
\begin{equation}
S_{\mathrm{new}} = \mathrm{logistic} (\mathrm{logit}(S_{\mathrm{current}}) - 1 )
\end{equation}
The mechanism of the update of tie strength caused by cultural
acceptance or rejection is illustrated in Figure \ref{updating}. If the tie
strength falls below 0.01, the tie is considered insignificant and is
removed from the social network.

\subsection{Initial Network Structures}

We set the network structures within and between merging firms so that
there are substantially more within-firm ties than between-firm ties
at the beginning of each simulation. The number of ties within each
merging firm is 490. Since the number of individuals in each firm is
50, the network density of the firm is 490/(50*49) = 0.2. The number
of ties from one merging firm to the other (that is, A$\to$B or
B$\to$A) is 50 for each direction. All tie strengths of those
connections are initialized using random numbers drawn from a uniform
distribution between 0 and 1.

In our computational experiments, we set two experimental parameters
that control topological characteristics of the initial social network
among individuals. One is what we call the {\em within-firm
  concentration}, denoted by variable $w$. This parameter determines
the probability for each individual to be selected as an information
source of a within-firm tie. It is mathematically defined as
\begin{equation}
  P_w(i) \sim ( i / n )^w \quad \quad ( i = 1, 2, \ldots , n ) ,
\end{equation}
where $i$ is the ID of the individual within a firm, $n$ the firm size
($n$ = 50 in our simulations), and $P_w(i)$ the probability for
individual $i$ to be selected as an information source when
within-firm ties are initially created. When $w = 0$, within-firm ties
are uniformly distributed within the firm so that the organizational
structure of the firm is ``flat''. For larger $w$, the within-firm
information sources are more concentrated on a small number of
individuals with greater IDs, which represent a highly centralized
organizational structure of the firm, such as that with a one-man
CEO. In our model, we used $w$ = 1, 3, 5, 10, 20, and 30.

\begin{figure}
\centering
Within-firm concentration\\~\\
\includegraphics[width=0.9\textwidth]{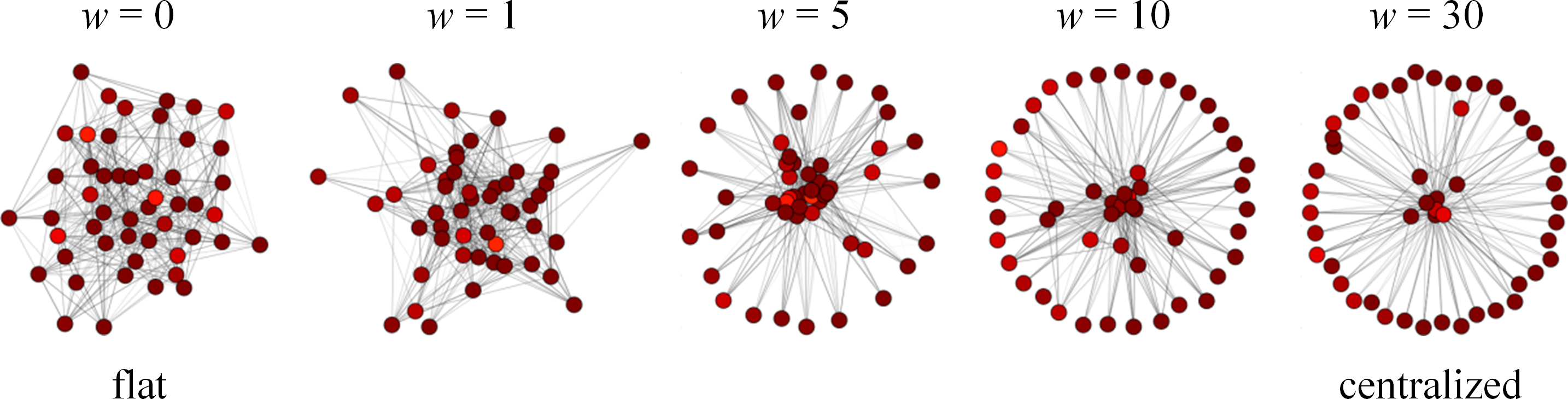}\\
~\\~\\
Between-firm concentration\\~\\
\includegraphics[width=0.9\textwidth]{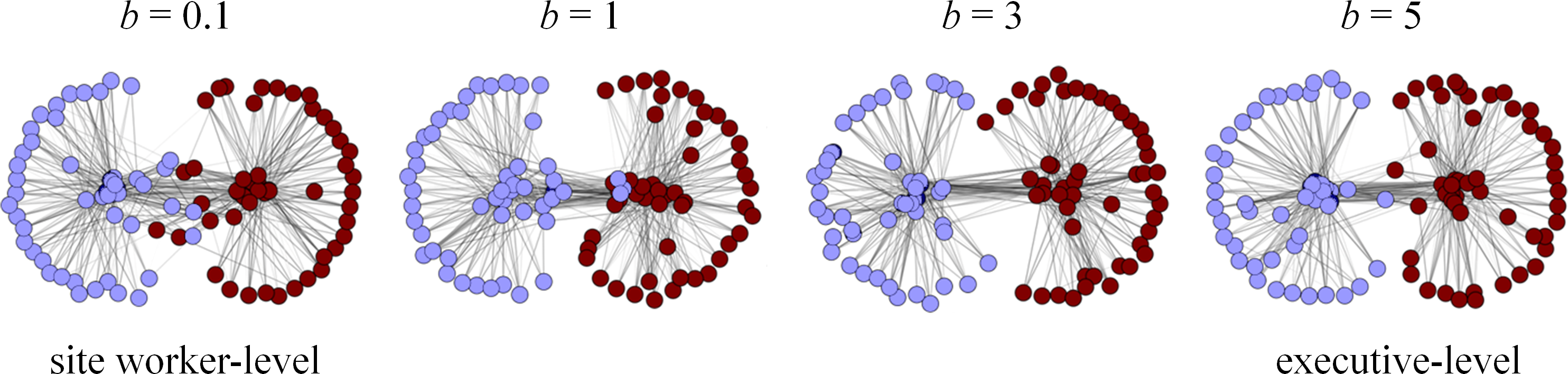}
\caption{Illustrations of within-firm and between-firm
  concentrations. The directions of ties are omitted in this figure
  for clarity (from \cite{lit35}).}
\label{w-b}
\end{figure}

The other experimental parameter is what we call the {\em between-firm
  concentration}, denoted by variable $b$. This parameter determines
the probability for each individual to be selected as either an origin
or a destination of a between-firm tie. It is mathematically defined
as
\begin{equation}
P_b(i) \sim c_i^b \quad \quad ( i = 1, 2, \ldots, n ) ,
\end{equation}
where $i$ and $n$ are the same as in the previous formula, $c_i$ the
within-firm closeness centrality of individual $i$, and $P_b(i)$ the
probability for individual $i$ to be selected as a connecting person,
either as origin or destination, when between-firm ties are created,
which is done only after all the within-firm ties have been
created. When $b = 0$, between-firm ties randomly connect individuals
across firms, regardless of their social positions. For larger $b$,
the between-firm ties are more concentrated on a small number of
individuals with higher centralities that represent the formation of
top-level (only) inter-firm communication channels. In our model, we
used $b$ = 0.1, 0.5, 1, 3, and 5. Figure \ref{w-b} illustrates images
of within-firm and between-firm concentrations.

Note that the above two parameters affect only the initial social
network structure. As cultural integration progresses, the network
topologies will change dynamically in our simulations.

\subsection{Outcome Measurements and Results}

As a primary dependent variable of our computational experiments, we
measure the average cultural distance between individuals who used to
belong to different pre-merger firms and who still remain in the
largest connected component of the social network. If the average
cultural distance decreases from its initial value, cultural
integration proceeds among individuals in the merged firm.

Likewise, we use three measures of the consequences of cultural
integration: turnover, interpersonal conflict, and organizational
communication ineffectiveness. All the measures should influence
overall firm performance. 

Turnover is measured by the number of individuals in the simulations
who do not stay in the largest connected component of the social
network. In our model, if an individual terminates all ties with his
neighbors, he is considered to have left the merged firm.

Interpersonal conflict is calculated as the cultural distance across a
social tie between two individuals, multiplied by their tie
strength. This quantity is summed up for all the tied pairs of
individuals within the largest connected component. Since tie strength
can be considered to represent communication frequency
\cite{granovetter1973}, individuals who are strongly tied to neighbors
with different perceptions of corporate culture would often encounter
greater communication conflict in the workplace.

Lastly, organizational communication ineffectiveness is calculated by
the cultural distance across a social tie between two individuals,
multiplied by the edge betweenness of the tie between them. This
quantity is, again, summed up for all the tied pairs of individuals
within the largest connected component. Edge betweenness is defined as
the number of geodesics (shortest paths) going through an edge
\cite{wasserman-faust1994}. If a tie with high edge betweenness is
filled with cultural conflict, most communication between individuals
in a firm would be conflicted. As a result, information and knowledge
transfer in the firm would be delayed or impeded.

We implemented the simulation model and analysis tools by using Python
with NetworkX. The program codes of the model are available from the
authors upon request. We set 200 time steps in one simulation. We ran
50 simulations for each experimental condition and conducted
statistical analysis of the generated simulation results.

\begin{figure}[t]
\centering
\includegraphics[width=0.7\textwidth]{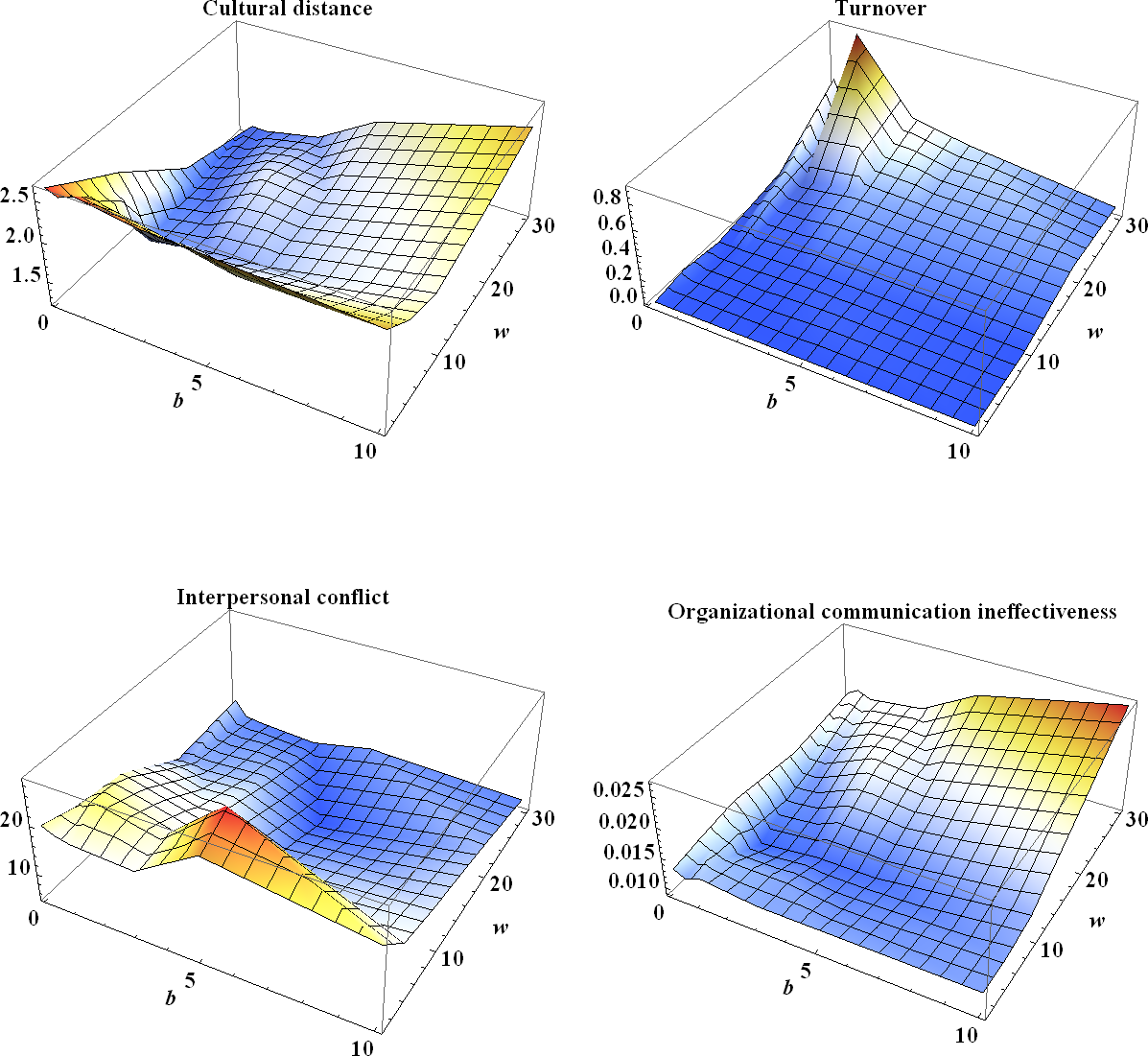}
\caption{Cultural distance and organizational dysfunctions by
  within-firm and between-firm concentrations obtained through
  simulations of our adaptive network model of corporate merger (from \cite{lit35}).}
\label{yamanoi-results}
\end{figure}

Figure \ref{yamanoi-results} plots the results showing the effects of
within-firm and between-firm concentrations. The highest level of
cultural integration is achieved when social ties are more centralized
within each merging firm and the social ties between the merging firms
are less concentrated on central individuals. Additionally, the
results show that within-firm and between-firm network structures
significantly affect individual turnover, interpersonal conflict and
organizational communication ineffectiveness, and that these three
outcome measurements do not vary in tandem. The most turnovers were
observed when within-firm concentration was high while between-firm
concentration was low, which is the same condition as that promoted
cultural integration. Interpersonal conflict was highest when
within-firm concentration was low, without much interaction with
between-firm concentration. Organizational communication
ineffectiveness was highest when both within- and between-firm
concentrations were high. For more detailed discussion of the results,
see \cite{lit35}.

Note that those findings were all the outcomes of the adaptive changes
of social ties in our model. Results would be different if the social
network structure was fixed like in other, more typical opinion
spreading network models.

\section{Conclusions}

As briefly reviewed above, the co-evolution of network states and
topologies is an emerging research topic that has great potential and
applicability to many real-world complex systems. It combines two
separate dynamics, i.e., dynamic state changes {\em on} a network and
topological transformations {\em of} a network, into a single picture
that will allow one to better understand and represent the nature of
evolving complex systems, possibly leading to new properties that were
not discovered before.

The application areas of adaptive networks are now expanding to
various disciplines, not only social sciences and operations research
(as demonstrated by a few examples in this paper) but also biology,
ecology and physical sciences. The key challenges in adaptive network
research include (1) how to generate meaningful dynamical models from
large-scale temporal network data, and (2) how to mathematically
analyze the dynamics of adaptive networks in which the time scales of
state changes and topological transformations are inseparable. We hope
that the work reviewed in this paper helps indicate the future
directions of this exciting field of study.

This material is based upon work supported by the US National Science
Foundation under Grant No. 1027752. The development of OpNetSim was
supported by the Canadian Government contract W7714-125419.

\bibliographystyle{elsarticle-num}

\end{document}